# Edge interventions can mitigate demographic and prestige disparities in the computer science coauthorship network

KATE BARNES, University of Colorado Boulder, USA
MIA ELLIS-EINHORN, Haverford College, USA
CAROLINA CHAVEZ, University of Colorado Boulder, USA
NAYERA HASAN, Haverford College, USA
MOHAMMAD FANOUS, Haverford College, USA
BLAIR D. SULLIVAN, The University of Utah, USA
SORELLE A. FRIEDLER, Haverford College, USA
AARON CLAUSET, University of Colorado Boulder, USA

Social factors such as demographic traits and institutional prestige structure the creation and dissemination of ideas in academic publishing. Such effects can be observed in how central or peripheral a researcher is within their field's coauthorship network. Here we investigate inequities in network centrality in a hand-collected data set of 5,670 U.S.-based faculty employed in Ph.D.-granting computer science departments and their DBLP coauthorship connections. We introduce novel demographic labeling algorithms that combine name- and perception-based labels, and show that these algorithms have high accuracy relative to self-reported demographic labels. We find that women, minoritized races, and faculty without parents employed in tenure-track academic positions tend to be less central in the computer science coauthorship network, implying worse access to and ability to spread information. Coauthorship centrality is also highly correlated with prestige, such that faculty at top-ranked departments tend to occupy positions in the network's core, while those at low-ranked departments tend to occupy positions in the periphery. We show that these disparities can be mitigated using simulated edge interventions, interpreted as facilitated scientific collaborations. The intervention increases the centrality of target individuals, chosen independently of the network structure, by linking them with researchers located at top-ranked institutions. When applied to scholars during their Ph.D., the intervention also improves the predicted institutional rank of their faculty job. This work uncovers social inequities in order to address them. By selecting scholars for intervention based on their demographics and institutional prestige, we can improve their centrality in the coauthorship network, potentially improving job placement and long-term academic success.

---

Authors' Contact Information: Kate Barnes, kathryn.barnes@colorado.edu, University of Colorado Boulder, Boulder, Colorado, USA; Mia Ellis-Einhorn, melliseinh@gmail.com, Haverford College, Haverford, Pennsylvania, USA; Carolina Chavez, carolina.chavez@colorado.edu, University of Colorado Boulder, Boulder, Colorado, USA; Nayera Hasan, nhasan1@haverford.edu, Haverford College, Haverford, Pennsylvania, USA; Mohammad Fanous, mfanous@haverford.edu, Haverford College, Haverford, Pennsylvania, USA; Blair D. Sullivan, sullivan@cs.utah.edu, The University of Utah, Salt Lake City, Utah, USA; Sorelle A. Friedler, sorelle@cs.haverford.edu, Haverford College, Haverford, Pennsylvania, USA; Aaron Clauset, aaron.clauset@colorado.edu, University of Colorado Boulder, Boulder, Colorado, USA.

---

## 1 Introduction

Prestige plays a central role in structuring a wide variety of outcomes in academic systems. For instance, the most prestigious 20% of U.S. universities train roughly 80% of all tenured and tenure-track faculty in every field at U.S.-based Ph.D.-granting institutions [Wapman et al. 2022]. Scientists at elite institutions publish more papers [Larivière et al. 2010; Samuel F Way et al. 2019], have larger research groups [Zhang et al. 2022], and receive more grant funding and citations to their papers [Larivière et al. 2010]. These prestige hierarchies are also remarkably stable over time [Bastedo and Bowman 2010; Lee et al. 2012], and effectively amplify the spread of ideas originating from elite institutions [Morgan, Economou, et al. 2018]. To adapt to this context a famous quote by Theodosius Dobzhansky [Dobzhansky 1973], it is reasonable to conclude that "*Nothing in academia makes sense except in the light of prestige hierarchies.*"

A strong prestige gradient is not by itself necessarily unfair, and broad statistical analyses suggest that steep prestige gradients in academia tend to reflect a combination of both genuine differences in merit and various non-meritocratic factors or processes [Clauset et al. 2015; LaBerge, Wapman, Morgan, et al. 2022; Morgan, LaBerge, et al. 2022]. Hence, we can ask whether it may be possible to intervene in some natural way to mitigate non-meritocratic disparities, including systemic devaluation of scientists from minority gender or racial groups or lower socioeconomic backgrounds [Casad et al. 2021; Dupree and Boykin 2021; Morgan, LaBerge, et al. 2022; Settles et al. 2021; Spoon, LaBerge, et al. 2023], or to improve the structural position or job prospects of promising individuals from non-elite institutions.

In highly collaborative fields like Computer Science, coauthorship on published papers is a key way that the community structures itself [Newman 2004; Sarigöl et al. 2014]. Facilitating novel scientific collaborations is thus a natural way to intervene in an academic system to potentially improve its fairness. Here, a more fair system would reduce the coupling of demographic disparities and centrality in the field's coauthorship network, equalizing social access to information and opportunities across protected groups. We choose to focus on centrality rather than typical indicators like citations, because we expect that increasing centrality will indirectly improve other indicators (centrality is highly correlated with citations [Sarigöl et al. 2014]) and confer other important but less measurable social benefits for individuals and the field. Practically, our intervention program could be implemented through targeted fellowship funding for collaborative projects with research groups at other institutions.

### 1.1 Contributions

We study how to make edge interventions into a real-world faculty coauthorship network to improve an individual's position in the system and their predicted academic job placement. Specifically, we make the following contributions.

**Computer science faculty coauthorship network.** We conduct a census of 5,670 computer science (CS) faculty in 178 Ph.D.-granting departments at U.S. institutions, collecting from public university websites each faculty member's name, current faculty rank, and current and doctoral institutions (Section 2). We create a novel CS faculty coauthorship network from this census using DBLP bibliographic data (Section 4), in which nodes (faculty) are also labeled with several measures of institutional prestige. We make a de-identified version of this high-quality social network and associated node-level attribute data publicly available for reuse by the community [Barnes et al. 2025].

**Demographic meta-labeling algorithms.** We introduce a novel name-and-perception-based algorithm to augment this real-world coauthorship network with faculty gender and racial demographic information, to facilitate the study of demographic-related disparities in computer science. These algorithms combine name-based demographic inference with perceived gender and race labels based on photos from public faculty websites, and are validated using self-reported demographic labels collected via a large-scale social survey of CS faculty (Section 3.3).

**Fairness findings.** We analyze the CS faculty coauthorship network and its associated institutional prestige, parental education status, gender, and racial labels, and we show that :

(1) Centrality within the CS coauthorship network correlates with (i) institutional prestige (Section 5), (ii) race, and (iii) parental education status (Section 5.3), such that Ph.D. students and faculty from highly ranked institutions, majority racial groups, and with parents employed in tenure or tenure-track positions at Ph.D. granting institutions tend to be more centrally located in the academic collaboration network than are individuals without these characteristics.
(2) Doctoral prestige and centrality within the coauthorship network are predictive of institutional prestige in faculty placement (Section 6.2).

**Intervention algorithm.** We introduce an intervention algorithm that selects—*without knowledge of the network itself*—a single edge to add to the network (i.e., creating a new coauthorship link, as by a facilitated new scientific collaboration). We show that when applied to Ph.D. students, this intervention increases both their closeness centrality and improves the predicted faculty job placement rank within the academic job market (Section 6).

## 2 Related Work

While institutional prestige has enormous influence on many aspects of faculty life, faculty experience can vary dramatically with social identity. A broad literature indicates that academic life can be strongly gendered, with negative consequences [Casad et al. 2021; Settles et al. 2021]. Women faculty leave tenured and tenure-track faculty positions at higher rates than men at all career stages; gendered attrition is higher at less prestigious institutions [Spoon, LaBerge, et al. 2023], and the shorter career lengths of women reduce their overall number of scientific contributions [J. Huang et al. 2020]. A key factor in such attrition is gendered devaluation, both formally in summative assessments and informally in department life [Spoon, Mendy, et al. 2024]. On the other hand, work on faculty hiring networks indicates that gender does not correlate with differences in job placement prestige in most fields [Wapman et al. 2022].

Racial minority faculty can also experience various forms of devaluation in the academy [Dupree and Boykin 2021; Settles et al. 2021; Spoon, Mendy, et al. 2024]. For instance, research shows they can be subjected to double standards in promotion evaluations [Masters-Waage et al. 2024] and to implicit biases during faculty hiring [White-Lewis 2020], and that their scholarship is funded at lower rates by grant-making agencies [Hoppe et al. 2019].

Race, gender and even parental education levels also influence the structure of scientific coauthorship networks [Bravo-Hermsdorff et al. 2019; Morgan, LaBerge, et al. 2022; Whittington et al. 2024]. For instance, women researchers often have fewer distinct collaborators than men [Zeng et al. 2016], which can reduce their effective productivity and prominence over their career [Li, Zhang, et al. 2022]. Additionally, researchers can exhibit gender or racial homophily, coauthoring with same-gender or same-race individuals at higher rates than expected by chance [Freeman and W. Huang 2015; Whittington et al. 2024] (but not always, see [Li, Zheng, et al. 2025], and our Section 5.2). Faculty with Ph.D. parents tend to have jobs at more prestigious institutions [Morgan, LaBerge, et al. 2022] and prestige is highly correlated with productivity and number of collaborators [Zhang et al. 2022]. Recent work suggests that women in computer science working on artificial intelligence tend to occupy less central positions in collaboration networks [Vlasceanu et al. 2022]. Collaboration network centrality is highly correlated with citations [Sarigöl et al. 2014; Yan and Ding 2009] implying practical disadvantages for such peripheral scholars. However, no studies have considered the broader question of how race, gender and/or parental education relate to centrality in computer science at large, or how strongly centrality in coauthorship networks correlates with prestige (see our Sections 3.1 and 5.3).

Such network-driven demographic disparities have received increasing algorithmic attention, with recent work often focused on interventions that aim to achieve fairness in social networks (for a survey, see [Saxena et al. 2024]). Definitions of fairness have largely focused on access to information propagated under some network flow model (often the independent cascade model), with the motivation that information and other resources (e.g., access to jobs) are often shared via social networks. Individual fairness definitions have aimed to maximize the minimum access of an individual to some information seeded in the network (e.g., [Fish et al. 2019]) or other individual notions of control, broadcast, or structural access in a network (e.g., [Bashardoust et al. 2023]), while group fairness definitions have aimed to equalize such information access across demographic groups represented as node attributes [Farnad et al. 2020; Stoica and Chaintreau 2019]. Intervention approaches have included changing who has access to information and direct intervention on the network structure via edge augmentation [Bashardoust et al. 2023; Becker et al. 2023; Bhaskara et al. 2024; Fish et al. 2019; Windham et al. 2024].

In this paper, we bring together these two lines of work; studying computer science faculty with a focus on demographics and prestige, and algorithmic fairness interventions in the context of real-world coauthorship networks.

## 3 Assembling a Census of CS Faculty

In the context of social analyses using potentially sensitive attributes, high-quality data is of great importance. To this end, we conducted a manual census of 5,670 computer science faculty at all 178 Ph.D. granting departments in the United States. The full list of universities, given in Appendix C.1, was compiled in previous research [Clauset et al. 2015] and includes all U.S. universities which grant Ph.D.'s in computer science. A de-identified version of our coauthorship network edge list and node metadata, which excludes personally identifiable information, is available on GitHub [Barnes et al. 2025].

*Data coders.* Seven undergraduate coders were hired during the 2023-24 academic year to conduct the census, and paid an hourly rate determined by institutional policy. Hired student coders were also given the opportunity to join the core research team, and one is an author on this paper.

*Coding procedure.* Coders were instructed to search institutions by name, navigate to the computer science department pages, and collect information for tenured and tenure-track faculty members. The hired coders were trained to follow a rigorous coding procedure (see Appendix A.1) to ensure consistency. Appendix A.2 provides the results of a subsequent error analysis assessing the accuracy of our census (error rate 2.75%). The error rate was calculated by double-collecting a random subset of the census and includes faculty who should not have been collected because they are not on the tenure track and inconsistent institution labels. In cases of disagreement, we retained the original coder's label, as the second label was used to assess reliability rather than to adjudicate individual cases.

### 3.1 Departmental prestige

In order to test the associations between prestige, demographics, and coauthorship in computer science (Section 6.2), we annotated the current and Ph.D. institutions of faculty in our census with rankings from three sources (see Appendix A.3): (i) the US News & World Report (USNWR) computer science graduate program ranking[1], (ii) CSRankings[2] and (iii) a prestige measure called "placement power" derived from faculty hiring networks [Wapman et al. 2022]. The Wapman et. al. (2022) ranking is highly correlated with CSRankings (Pearson's $r = 0.77$, $p = 8.5 \times 10^{-53}$) and USNWR (Pearson's $r = 0.88$, $p = 1.5 \times 10^{-33}$) but more accurately predicts faculty placement [Clauset et al. 2015].

## 3.2 Augmenting the census with demographic variables

The terms "race" and "gender", instead of "ethnicity" or "sex", are used throughout to reflect the socially constructed nature of the demographic categories we study. Mislabeling gender or race attributes can have harmful impacts. In research contexts these include misrepresenting population sizes and underlying patterns of marginalization. To mitigate potentially harmful errors in our demographic labeling of faculty in our census, we consider three sources of demographic information: perception, name-based inference, and survey self-identification. Race was categorized in alignment with 2024 U.S. Census Bureau categories [Bureau 2024], with the exception of the U.S. Census Asian category which we expand to include East Asian, South Asian (Indian / Indian subcontinent) and Southeast Asian.

*3.2.1 Perception of race and gender.* Perception data about race and gender was collected by the hired student coders based on publicly available photos, e.g., on faculty directories and faculty websitess. Race and gender characteristics that are socially perceived do not always align with individual's lived experience or self-identification. However, many patterns of marginalization are perpetuated on the basis of perceived traits, which constitute part of the social constructions. The full procedure for coding demographics and an error analysis (89% and 100% agreement on perceived race and gender respectively) are given in Appendices A.1 and A.2.

*3.2.2 Name-based inference of race and gender.* In addition to perception, we employed four name-based gender and five name-based race inference tools, using only those well-aligned with survey information (see Appendix A.4). Despite the intrinsic limitations of name-based demographic labeling, high probability labels indicate strong alignment between the given name and an individual's gender [Buskirk et al. 2023] and race attributes [Kozlowski et al. 2022], understood within the cultural consensus framework [Haslanger 2000].

*3.2.3 Survey self-identification of race and gender.* Our IRB-approved[3] survey received 820 responses, for a response rate of 15%. This rate is similar to those of other surveys of faculty [Morgan, LaBerge, et al. 2022; Morgan, Samuel F. Way, et al. 2021; Spoon, LaBerge, et al. 2023]. Of the respondents, 75% were men, 84% identified with a majority racial group, and 16% had a parent employed in tenure or tenure-track positions at a Ph.D. granting institution. Although surveys can be subject to various sampling biases [Elston 2021], they are the gold standard for demographic labeling [Lockhart et al. 2023]. The complete text of the survey questionnaire is given in Appendix A.5.

## 3.3 Meta-labeling algorithms for gender and race

With survey self-reports to validate against, we developed two meta-labeling algorithms that combine the perceived and name-based demographic labels for a more accurate estimate. Both our race and gender meta-labelers proceed by first applying name-based inference to all names and accepting only high-confidence labels (confidence greater than some per-method threshold). Names with assigned label confidence below threshold are then classified using perception labels. Given the high alignment between perception and self-reported data, the idea behind first using name-based inference followed by perception is that future researchers could use this generalized methodology to start with the more scalable name-based inference methods, and use the more expensive, hand-collected perception data only when necessary. See Appendix A.6 for a general schematic of the gender and race meta-labelers.

[1]https://www.usnews.com/best-graduate-schools/top-science-schools/computer-science-rankings
[2]https://csrankings.org/#/fromyear/2014/toyear/2024/index?all&us
[3]Approved by the Haverford College IRB dated June 17, 2024 under the title "Demographics and Faculty Co-Authorship Networks."

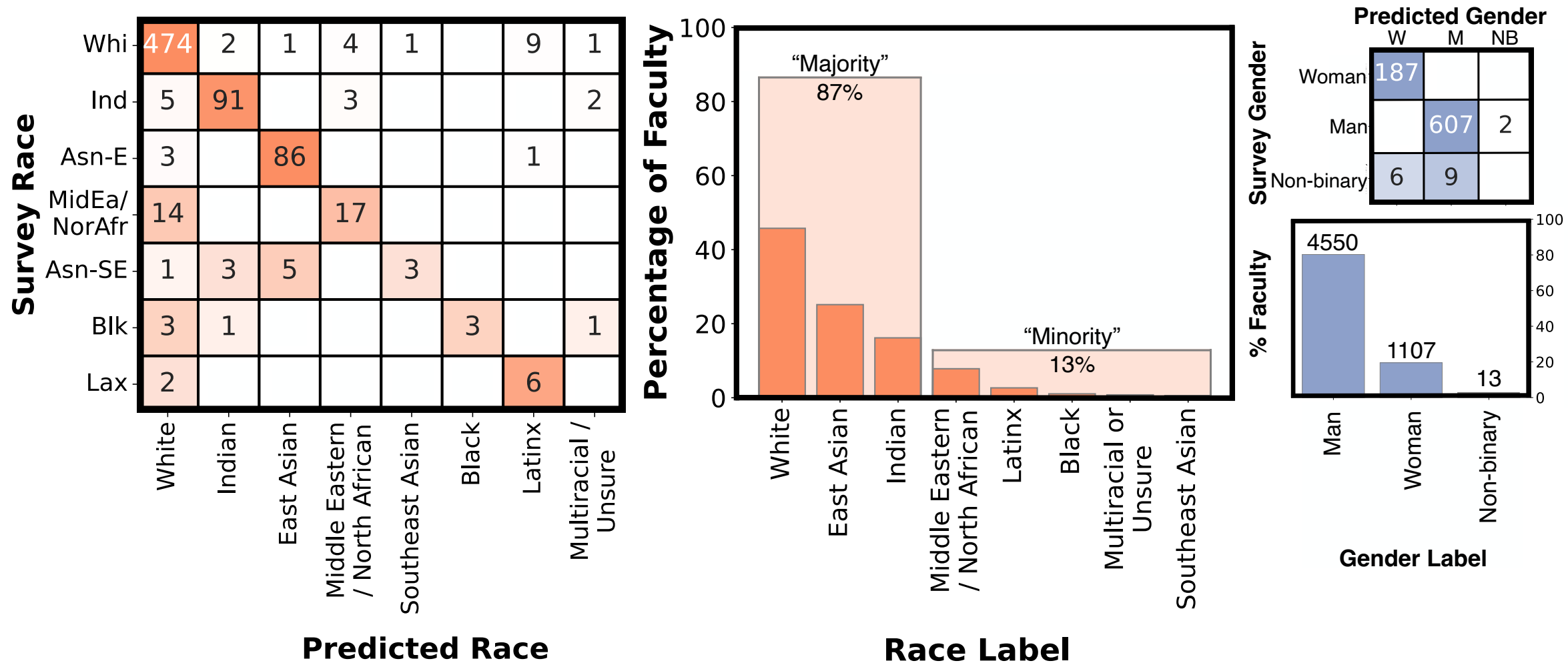

Fig. 1. Confusion matrices showing agreement between our meta-labeling algorithms and survey self-reports. Distributions of demographic meta-labels for all 5,670 computer science faculty in our census.

*3.3.1 Race meta-labeling.* The included inference methods for the race meta-labeler are perception, EthnicolrWiki, EthnicolrFlorida [Laohaprapanon et al. 2022], and Ethnea [Torvik and Agarwal 2016]. For each name-based method, we selected separate confidence thresholds for every race by using the lowest threshold that still minimized false positives. The order in which racial labels are applied is critical to minimizing false positives by the algorithm. Using a brute force search, we examine all permutations of the racial label order, and choose the ordering that maximized accuracy relative to self-reported labels from the survey data. The final race meta-labeler algorithm (see Algorithm 2 in Appendix A.7) correctly labeled 92% of (733/797) survey respondents who answered race related questions. This algorithm was validated using 5-fold cross-validation and through comparison with a different exhaustive methodology (see Appendix A.7).

*3.3.2 Gender meta-labeling.* We solely consider the NonQuamGender [Buskirk et al. 2023] inference method for the gender meta-labeler, since it out-performed the other name-based inference methods, and searched over possible thresholds which maximized the accuracy of the resulting label with respect to self-reported labels via the survey data. East Asian names, known to have lower gender signal [Buskirk et al. 2023], were disproportionately labeled incorrectly by NonQuam. Therefore, our gender meta-labeler requires higher confidence for labeling East Asian names. The final gender meta-labeler first assigns NonQuam estimates gendered with $p > 0.85$ for those names identified as East Asian by Ethnea, and then assigns NonQuam estimates to all other names gendered with $p > 0.75$. Names that did not reach these confidence thresholds were labeled by perception. Complete pseudocode for the gender meta-labeler is provided in Algorithm 3 in Appendix A.8. Overall, Algorithm 3 achieves an accuracy of 98% (794/811) on survey data.

*3.3.3 Results and validation.* We labeled all 5,670 individuals in our data using Algorithms 2 and 3. In agreement with previous studies of computer science faculty [LaBerge, Wapman, Clauset, et al. 2024], CS faculty in our study are primarily men, with close to 20% women and 13 non-binary scholars. Racially, CS faculty in our study are primarily White, East Asian and Indian scholars, with racial minorities making up 13% of the population combined (Figure 1). These demographic proportions are in close agreement with those reported in the 2022-23 CRA Taulbee Survey[4] and with our survey responses (92% race and 98% gender agreement). Note that we cannot infer tenure-track parental status from public data, and our results for this trait use only data for the 133 survey respondents with and 687 without tenure-track parents.

Our methods maximize alignment with self-reported labels in order to most appropriately label faculty in our census. However, the meta-labeling algorithms exclusively apply perception or name-based inference labels. These are social signals of race and gender, and therefore our final labels should be thought of as reflecting sociological formations over the population at large rather than the self-identification of particular individuals [Gautam et al. 2024; Weitman 1981].

## 4 Constructing CS Coauthorship Networks

Factors like institutional prestige and demographic traits structure social aspects of academic scholarship. In order to study these factors, we construct coauthorship networks from 3,652,370 journal and 3,563,465 conference publications authored by faculty in our census (see Appendix A.1 for details), as indexed by the DBLP Computer Science Bibliography.[5]

*Cumulative CS faculty coauthorship network.* Nodes in this network are faculty from our census, and each pair of nodes is connected by an undirected edge weighted with the number of their coauthored papers. The edges are additionally annotated with a dictionary indicating how many papers faculty members coauthored together in each year. The network has $n = 5,348$ nodes and $m = 35,551$ edges. Only 323 faculty from our census are not included in the network due to having no DBLP-indexed papers or no collaborations with other individuals in our census. Based on a manual examination of these excluded faculty, we believe these are largely researchers in interdisciplinary computing who publish in venues that are not indexed by DBLP, e.g., *Science* or *PNAS*, or recent hires from international institutions who may not have yet collaborated with other U.S. computer science faculty.

*Ph.D. coauthorship networks by year.* We constructed networks representing collaboration patterns during scholars' Ph.D. periods by filtering the cumulative collaboration network. These Ph.D. networks are used to assess how a graduating student's position in the network shapes their placement outcomes on the academic job market (Section 6.2). We only considered Ph.D. networks for the 2,041 *early career* faculty in our census, those whose year of first publication was 2010 or later. The Ph.D. periods of faculty who began publishing earlier than this would be poorly represented by our data which does not include retired faculty who may have been prominent during their Ph.D. periods.

For early career scholars, we approximate the DBLP collaboration network at the time of their Ph.D. by identifying their year of first publication, $t_1$, in the DBLP data and then filtering the cumulative coauthorship network to contain only edges for publications from years $t \leq t_1 + 5$. Thus, these networks represent subsets of the total DBLP publication data, including only those papers which were published before or during the cutoff year. We built one Ph.D. network for each year between 2015 and 2024. Then, for example, the Ph.D. periods of faculty whose year of first publication was $t_1 = 2010$ are represented by the 2015 Ph.D. network, and those with $t_1 = 2011$ are represented by the 2016 Ph.D. network, etc.

[4]See https://cra.org/resources/taulbee-survey/ and Appendix A.9.
[5]The DBLP online repository of publications from major computer science venues: https://dblp.org

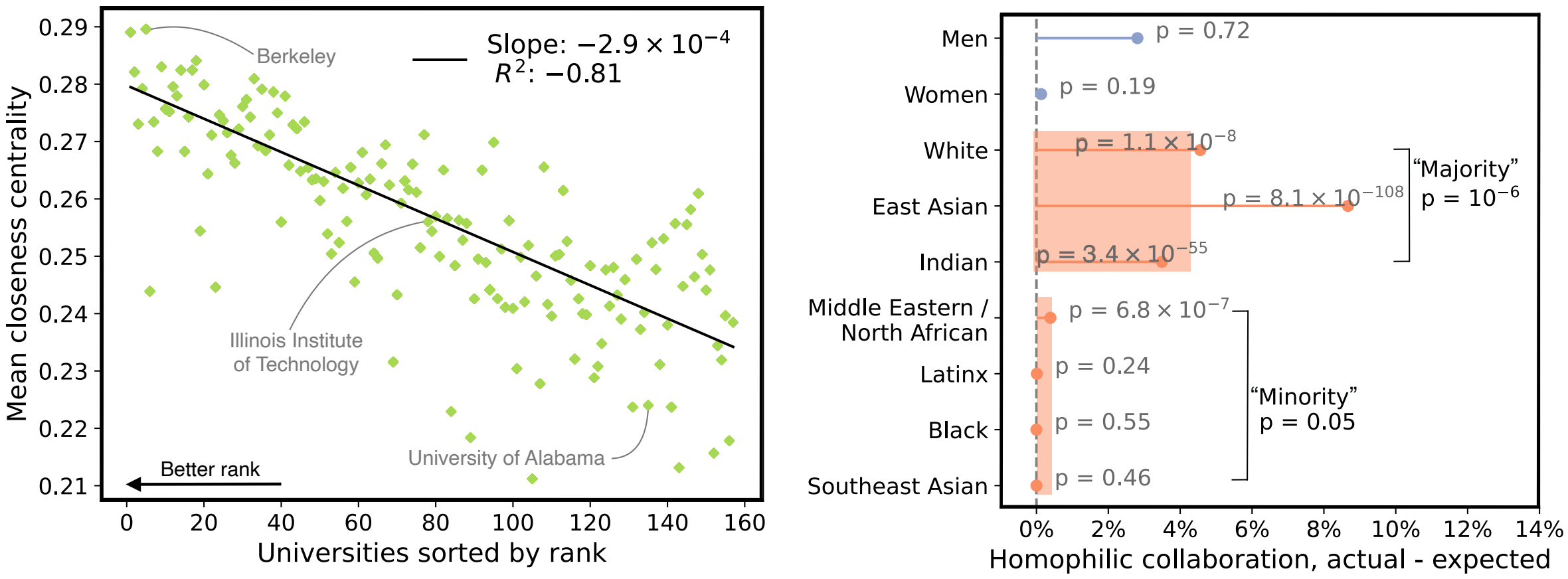

Fig. 2. (Left) University prestige (rank, using the placement power measure [Wapman et al. 2022]; smaller score is more prestigious) versus the mean closeness centrality of faculty at that institution. (Right) Difference between observed frequency of homophilic collaborations in the CS coauthorship network, relative to its expectation under a permutation test that keeps the network fixed while shuffling node labels; p-values indicate the expected fraction of permutations with greater homophily under a Gaussian approximation of the sampled distribution.

## 5 The Structure of Computer Science Collaboration Networks

Centrality measures characterize differences in node position within a network, and are commonly interpreted as proxies for a node's structural importance or social influence [Burt 2004; Wasserman and Faust 1994]. In scientific coauthorship networks, individuals with higher centrality can be expected to have greater or easier access to other parts of the network, and can act as bridges in receiving or distributing ideas. In contrast, lower centrality individuals have less or lower access to the same, a kind of structurally-mediated epistemic disparity. Centrality in coauthorship networks is also highly correlated with various other measures of academic success such as citation counts [Sarigöl et al. 2014; Yan and Ding 2009]. Here, we primarily focus on a node $v_i$'s closeness centrality $C(v_i)$, defined as,

$$C(v_i) = \frac{n-1}{\sum_{v_j \in V \setminus v_i} d(v_i, v_j)} \quad , \tag{1}$$

where $d(v_i, v_j)$ is the shortest-path (geodesic) distance between nodes $v_i$ and $v_j$ in a graph with $|V| = n$ vertices. Closeness centrality measures the inverse average path length to all other nodes in the network. Results for betweenness centrality, measuring how often a node falls on the shortest paths between other nodes, are analyzed in Appendix B and closely mirror those for closeness centrality.

Closeness and betweenness centralities were calculated for all 5,348 individuals in the cumulative coauthorship network, and for the 2,041 early career scholars in each Ph.D. networks. Centrality measures are highly correlated with node degree, which denotes the number of connections (collaborators) in the network. Hence, for comparison, we also calculated nodes' normalized degrees to test the residual influence of faculty centrality, while accounting for their simple number of collaborators.

### 5.1 Faculty centrality and institutional prestige

We find a strong correlation between a faculty member's centrality in the CS coauthorship network and the prestige of their current institution—a finding not previously reported in studies of scientific coauthorship [Vlasceanu et al. 2022; Zhang et al. 2022]. Faculty in top-ranked CS departments are highly central in the network (higher closeness scores), whereas faculty at bottom ranked institutions are systematically more peripheral. This pattern appears as a strong linear relationship between closeness centrality and prestige rank ($R^2 = -0.81$, Figure 2), showing that scientific collaboration also aligns with the prestige hierarchy in computer science, with densely interconnected coauthorships among computer scientists at the most prestigious institutions, and more diffuse coauthorships among those at less prestigious institutions. Such a strong core-periphery pattern may further reinforce the way prestige-structured faculty hiring networks can skew the spread of ideas in computer science, undervaluing good ideas that originate outside the core [Morgan, Economou, et al. 2018]. This pattern is not sensitive to the particular measure of institutional prestige (see results for CSRankings and USNWR university rankings Appendix B).

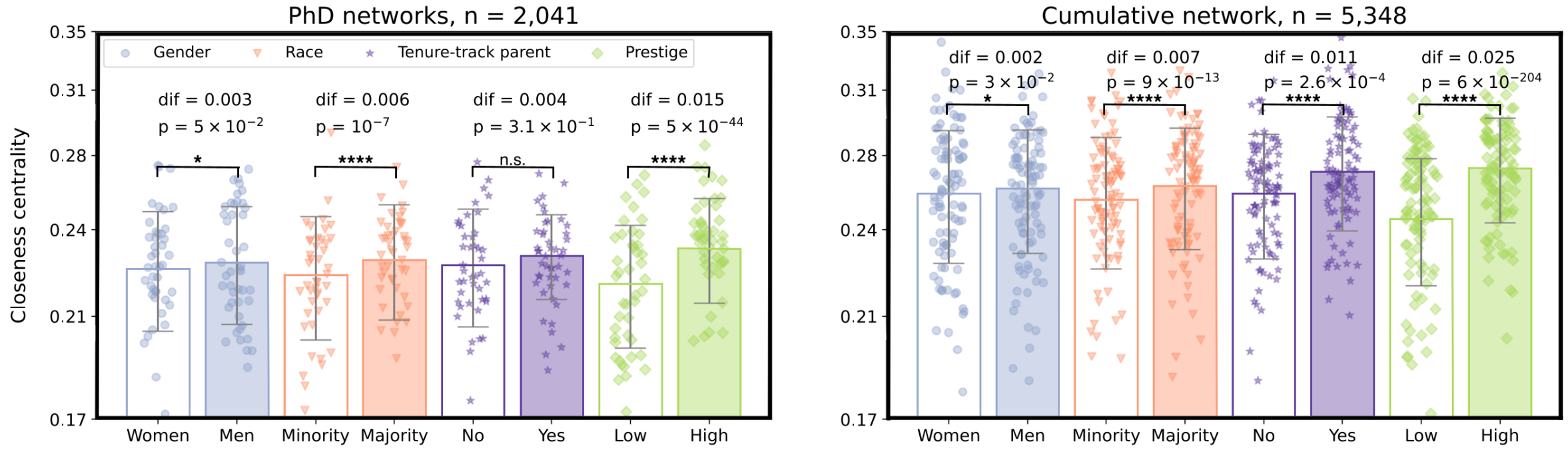

Fig. 3. Group disparities in network position, showing significant differences in closeness centrality (pooled t-tests) between (i) women and men, (ii) racial minority and majority groups, (iii) survey respondents who did and did not have a parent in a tenure-track academic position (Ph.D. $n = 310$, cumulative $n = 820$); and (iv) low and high prestige institutions (split at the median rank).

## 5.2 Homophily in coauthorship

Homophily is an empirical pattern in which members of the same social group are more likely to share social links with each other than with members of other groups. We measure the prevalence of race- and gender-based collaboration homophily in the CS coauthorship network by comparing the observed frequency of homophilic collaborations (same race, or same gender) to the expected frequency given random mixing with the observed demographic distributions in our network. We calculate the expected frequency using a permutation model, in which the coauthorship network structure is fixed, and we measure the frequency of edges connecting nodes with the same label under random permutations of the demographic labels over the nodes. These comparisons between the observed and expected frequencies of homophilic collaborations are shown in Figure 2; $p$-values for these comparisons are calculated relative to a Gaussian approximation, parameterized by 1000 samples, of frequencies under the permutation model. Note that we observe no correlation between demographics and institutional prestige, so this analysis is unlikely to be confounded by prestige. The permutation model furthermore controls for differences in node degree that may influence homophily.

We find strong evidence of racial homophily among majority race groups (White, East Asian and Indian) in the CS coauthorship network (Figure 2). In contrast, the racial homophily rates among most minority race groups (Latinx, Black, Southeast Asian) are not significantly different from their expected values, although we do observe a significant difference for the largest minoritized group (Middle Eastern / North African, $p = 6.8 \times 10^{-7}$). The significant differences for majority racial groups and non-significant differences for most minorities may reflect a lack of statistical power that hides homophily among individuals within these minority groups because of their overall low representation, or it could indicate genuine non-significance due to, e.g., limited practical options for homophilic coauthorships.

For both men and women, the observed rates of homophilic collaboration were not significantly different from expectations. This negative result contrasts with some previous results suggesting that men, but not women, exhibit homophily in their scientific collaborations [Kwiek and Roszka 2021; Löther and Freund 2024]. This difference may indicate that correctly measuring homophily in coauthorship networks requires appropriately controlling for network structure, as in our permutation test, or it may indicate collaboration patterns specific to computer science, e.g., in how preferences for homophily or diversity are socialized [Li, Zheng, et al. 2025].

## 5.3 Disparities in network position

Women, scholars with minoritized race identities, and those from low-ranked institutions are less central in the CS coauthorship network than are men, scholars in majority race groups, and those from high-ranked institutions (Figure 3). Additionally faculty whose survey responses indicated that they had a parent employed in a tenure-track academic position are more centrally located than those without. These disparities in coauthorship centrality have not been observed before, and while the prestige disparity may partially reflect merit, previous literature shows that prestige effects in academia are not purely meritocratic—e.g., blinding institution names in review increases success rates for awardees from lower-ranked institutions [Hultgren et al. 2024]. Parental education is not explored in subsequent sections as these data were only available for survey respondents, however this disparity suggests another feature along which a targeted invervention could be conducted. The disparities in Figure 3 are replicated using betweenness centrality and degree in Appendix B. We found no correlation between our demographic variables and institution rank for either the Ph.D. or current institutions of faculty in our census, suggesting that the demographic disparities in closeness centrality presented in Figure 3 cannot be explained by the institutional career trajectories of minoritized scholars.

In addition to demographic disparities in the network centrality, we found that individuals with minoritized race identities had on average fewer collaborators ($p = 1.2 \times 10^{-4}$) and fewer papers ($p = 4.6 \times 10^{-5}$) than individuals from the majority. In pairwise comparisons, these disparities also held for gender, replicating previous results [Zeng et al. 2016]. However, controlling researchers' number of years since first publication (academic age), this result was no longer significant. This indicates that the apparent gender difference in degree and productivity is driven by a proportionally smaller number of advanced-career women scholars. This is also reflected in previous research which has shown that the predominancy of women scholars in early-career stages is largely due to increased hiring of women in recent years [LaBerge, Wapman, Clauset, et al. 2024], and to a lesser degree, by higher rates of attrition among women than men in U.S. faculty [LaBerge, Wapman, Clauset, et al. 2024; Spoon, LaBerge, et al. 2023].

# 6 Intervening to Address Demographic and Prestige Disparities

We now introduce a simple network intervention to help mitigate observed disparities in network centrality. This intervention adds one single edge to the coauthorship network by pairing a target individual with certain characteristics with a sponsor individual, e.g., by providing by a research fellowship for the target to spend a semester in the research group of the sponsor. We focus on the single-edge intervention as a minimal test case, which can be compared to baseline and theoretical maximum impacts of the intervention. In practice, a new collaboration may result in multiple new coauthorship edges, and thus our simulation should underestimates the potential gain of such a program.

First, we show that our intervention increases targets' centralities in the coauthorship network, improving their relative status in the research community. The centrality of sponsor individuals also increases, though to a much lesser degree. Second, we show that when applied to scholars in their Ph.D. periods, the resulting improvements in centrality also improve their predicted placement outcomes on the faculty job market.

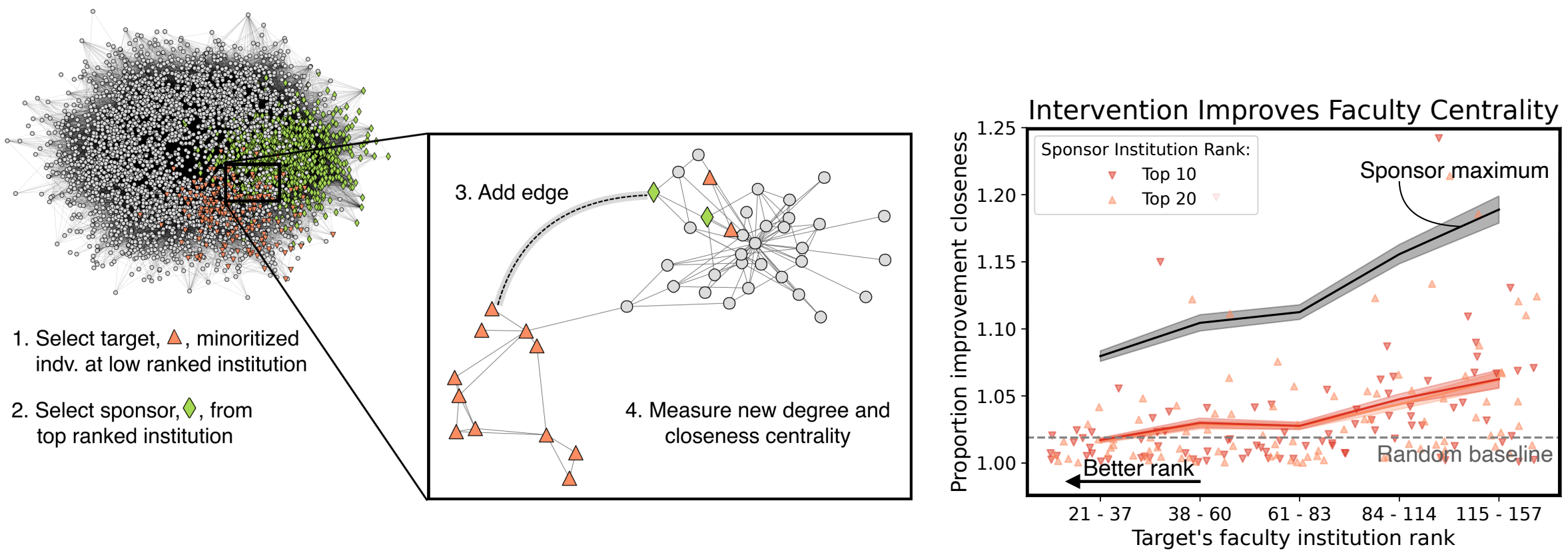

Fig. 4. (Left) Schematic of the proposed intervention to improve the centrality of target individuals. (Right) This simulated intervention results in increases in the closeness centrality of target individuals with minoritized race identities at low prestige current institutions (rank using placement power [Wapman et al. 2022]). Our interventions (red lines) are compared to greedy selection of sponsor who would maximally improve target's closeness (black line) and to a random baseline.

## 6.1 Suggested collaborations improve target centrality

In general, a peripheral node's centrality can be improved by linking it with a highly central node in the network. An intervention that specifically targets individuals with low centrality is likely infeasible in practice because it would requires constant access to and analysis of the full collaboration network. Instead, our simulated intervention targets individuals with minoritized race identities who are currently employed at lower-ranked institutions, whom our results suggest are statistically likely to be peripherally located in the coauthorship network. These individual characteristics define the greatest disparities in centrality for which we have complete data, but in practice any low-centrality group from Fig. 3 could be used for selection. We will show that this principled way of selecting target nodes results in greater improvements in centrality than would be expected from adding a random edge to the network.

Institutions are binned into five groups (ranks from 21–37, 38–60, 61–83, 84–114 or 115–157) and we define target groups for each bin as the set of minority race faculty at that bin's institutions. We consider two sponsor definitions: (i) all faculty at top-10 ranked institutions and (ii) all faculty at top-20 ranked institutions (both statistically likely to be centrally located because of their high prestige affiliations). For every individual in a target group, we apply the simulated intervention shown in Fig. 4. First, we select a random individual from the sponsor group and add an edge connecting the current target to the selected sponsor. This selection can be thought of as the funded fellow choosing someone from a list of participating sponsors. The sponsor groups are quite large (over 1000 individuals) relative to the target group (250 individuals), making it feasible for target scholars to select sponsors who are topically aligned with their research interests. Finally, we measure the post-intervention closeness centrality of target and sponsor.

We find that the simulated intervention results in a positive improvement in the centrality of every individual, for both targets and sponsors (Fig. 4), although the improvement is far larger for the target. Moreover, the positive slope in Fig. 4 shows that the network intervention has a greater effect for scholars at lower ranked institutions.

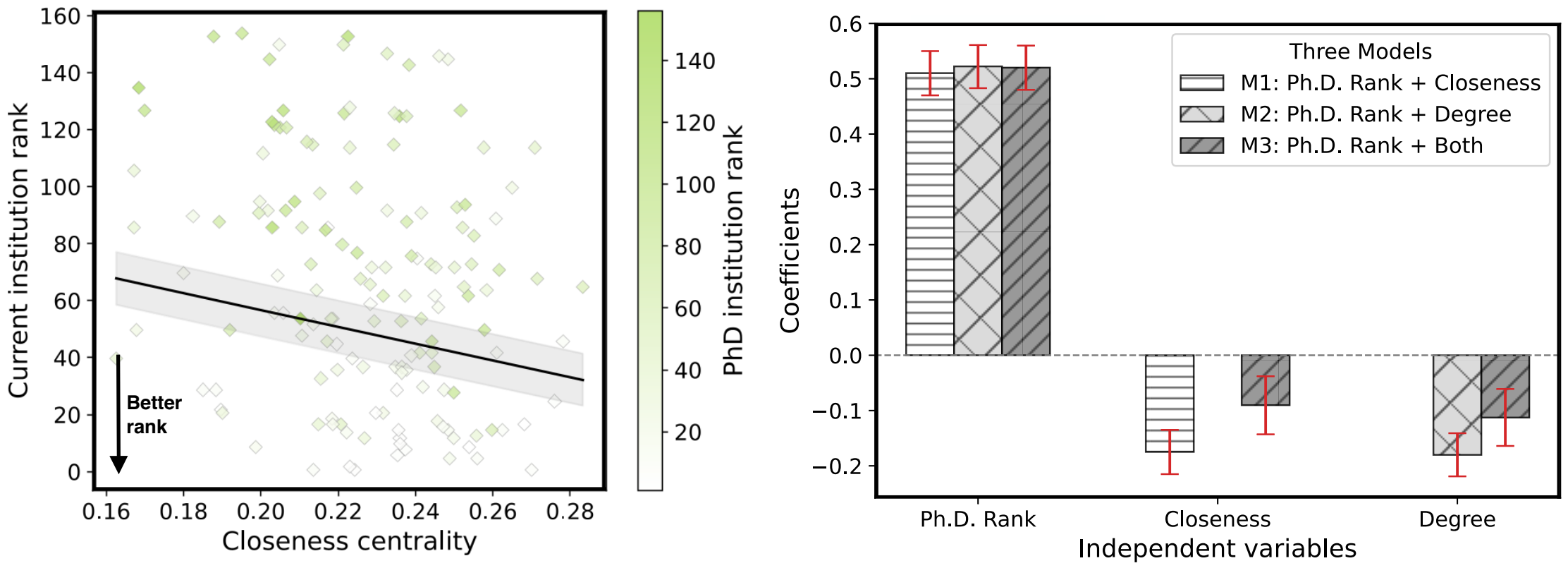

Fig. 5. Ph.D. closeness centrality and Ph.D. institution rank predict faculty current institution ranks. Model M1 line of best fit through a random subset of 150 points from fit data (Left) and standardized coefficients (Right) of three linear regression models.

Defining sponsor individuals as those from top-10 or top-20 ranked institutions makes little difference to the results.

We compare these effects with the maximum improvement possible and a random expected baseline change in centrality from a single edge intervention, showing that our intervention is, on average, substantially better than the random baseline, and in some cases competitive with the sponsor maximum. The random baseline was calculated as the average improvement in centrality from adding one edge between a randomly selected pair of unconnected individuals anywhere in the network. The greedy maximum centrality improvement in Fig. 4 is calculated by connecting each target individual with the best of all possible sponsors (maximum increase in centrality).

## 6.2 Predicting job placement based on Ph.D. prestige, centrality and degree

Beyond the inherent benefits of increasing network centrality, such as increased access to information and opportunities, we conjecture that centrality plays a practical role in improving individuals' career prospects. When an early career scholar enters the job market, it matters how many professional connections they have and to whom they connect. We test this conjecture by predicting the rank of the hiring institution for each faculty in the census via a regression on rank of their Ph.D. institution, network degree (number of coauthors), and closeness centrality. These models are trained on the data from 2,041 early career scholars in our census, using their Ph.D. network degree and closeness centralities (Section 4). Strictly interpreted, this analysis assumes that the current institutions in our census are the same institutions that first hired that particular faculty. Given that most faculty remain at their first university job [Samuel F Way et al. 2019], we find this to be a reasonable assumption.

We find that while Ph.D. prestige is the most important factor in predicting placement, Ph.D. network degree and centrality both have a significant effect (Fig. 5). The first model (M1) shows how current institution rank depends on Ph.D. institution rank and Ph.D. network closeness centrality. The positive coefficient for Ph.D. rank shows that individuals from low-ranked Ph.D. institutions are likely to be hired at low-ranked institutions, and vice versa for high rank. The negative coefficient for closeness centrality shows that more central nodes tend to be placed at better institutions. Although Ph.D. rank is more than twice as important to the model, closeness is still a significant predictor of current institution rank. The second model (M2), which predicts placement prestige using Ph.D. prestige and degree, replicates these results. Since degree and closeness are calculated within the Ph.D. networks, this demonstrates that individuals' coauthorship positions during their Ph.D. periods matters to their later career outcomes.

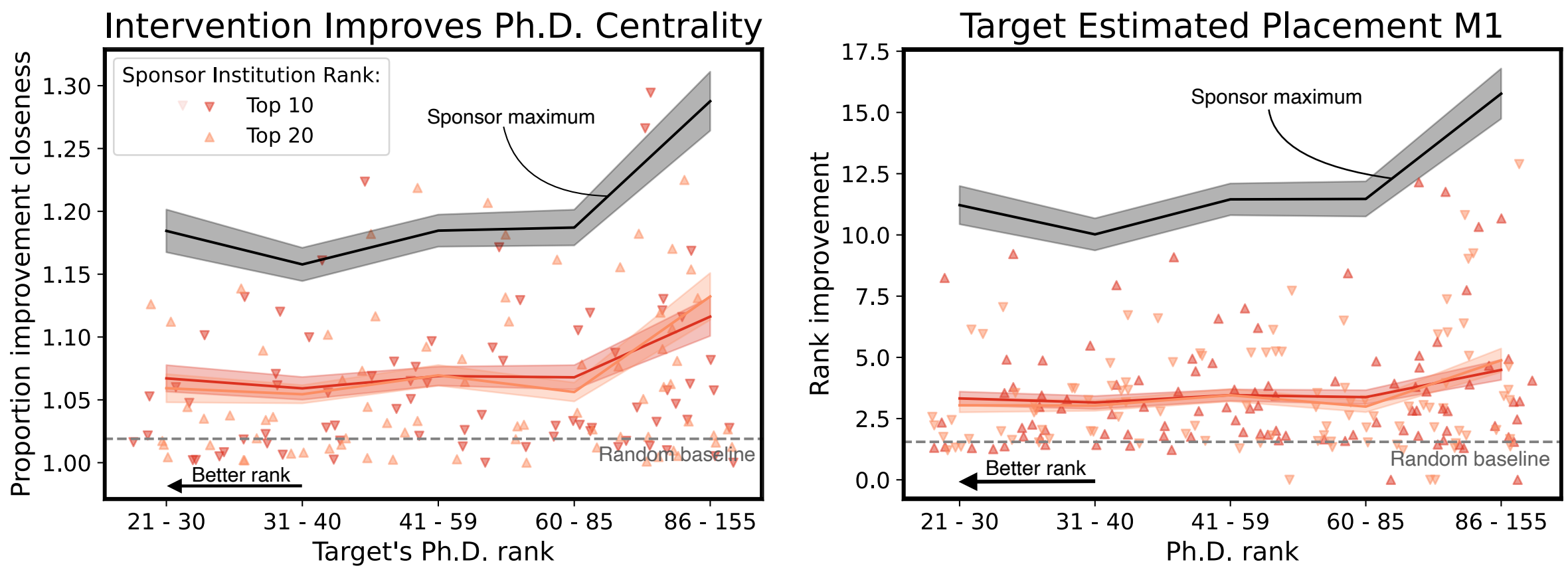

Fig. 6. Simulated intervention adds one edge between sponsor from a top-ranked institution and a racially minoritized target from low-ranked Ph.D. institution (placement power rank [Wapman et al. 2022]). The intervention increases the centrality (Left) of target individuals and improves the estimated rank (Right) of the institution at which the place for a faculty job. Comparisons show a greedy selection of the sponsor that maximally improves the target's closeness (black line, averaged over the bin) and to the random baseline.

Together, the M1 and M2 models show that individuals' position and number of connections within the coauthorship network at the time of their Ph.D. are individually predictive of placement outcomes, above and beyond the predictive value of Ph.D. institution prestige. Closeness is highly correlated with network degree, and this co-linearity is evident in the decreased importance of both factors in the third model (M3), which predicts current institution rank based on Ph.D. rank, network degree and closeness centrality. However, while degree has a mediating effect on closeness in M3, both factors remain significant predictors of current institution rank, of comparable importance. The M3 results indicate that when predicting placement in the faculty hiring market, in addition to the number of coauthors (degree), it matters *which* specific coauthors (closeness) a researcher has.

The idea that it matters who you collaborate with is corroborated by the intervention results in Fig. 4, where pairing targets from minoritized racial groups at low-prestige institutions with sponsor individuals at high-prestige institutions is substantially more effective for improving targets' centralities than pairing randomly chosen individuals.

## 6.3 Improving Ph.D. centrality and predicted rank of placement institution

The correlation between centrality and placement institution rank (Fig. 5) suggests that intervening to improve the centrality of early career scholars may have the material benefit of improving the rank of institution they are hired at on the academic job market. We use the regression model from the previous section to illustrate potential material benefits of our intervention.

We test this possibility with our second intervention experiment applied to the Ph.D. coauthorship networks. This intervention again uses minoritized race and lower Ph.D. institution rank to define the target populations, and we again bin institutions by their Ph.D. rank, ensuring an equal number of Ph.D. scholars in each of five bins. For each target individual in a given bin, a random sponsor is selected from the sponsor group, with the additional restriction that sponsors must have a year of first publication at least five years prior to the target individual's year of first publication (which ensures that they are more senior than the target individual). We then add the new coauthorship edge and calculate the post-intervention closeness centrality of both target and sponsor in the Ph.D. coauthorship network.

The improvement in centrality resulting from this intervention is compared to a random baseline and greedy maximum, using the same strategy as our first intervention experiment. Notably, the proportional increase in target Ph.D. centrality (Fig. 6) is greater than the improvement found in the first experiment. Ph.D. centrality is calculated using fewer coauthorships—only the first five years of the target individual's publication record—meaning that one added coauthorship has a bigger impact on their network centrality at this time than it does in their later career.

Improving one's centrality in the academic network has inherent benefits, such as increased ability to distribute one's scientific ideas, and may improve the rank of the institution a scholar is hired at. We estimated the improved ranks of target individual's placement institutions using M1 with the updated closeness centrality of target individuals post-intervention. These estimates place target scholars at a higher ranked institution than their current institution in our census (Fig. 6). Scholars from the lowest ranked universities (86-155) experience an estimated improvement of 4.4 in the rank of their resulting university job placement. Although the simulated intervention suggests a causal relationship between targets' Ph.D. centrality and placement institution ranks, our results do not empirically assert such a relationship. In order to make that claim, the intervention would need to be carried out in practice, e.g., implemented via a fellowship program, presenting a valuable direction for future research.

## 7 Discussion

These results expand on previous studies showing that differences in institutional prestige drive many aspects of academic success [LaBerge, Wapman, Morgan, et al. 2022; Wapman et al. 2022; Samuel F Way et al. 2019; Zhang et al. 2022]. In parallel with findings from previous research [Morgan, Economou, et al. 2018], which showed that prestige amplifies the spread of scientific ideas originating at top universities, our results highlight socially disadvantaged groups that are epistemically dis-empowered in receiving and distributing scientific knowledge. Furthermore, for Ph.D. scholars, these network attributes have practical consequences in determining their placement institutions on the academic job market (Section 6.2).

In order to improve the fairness of this system (i.e. mitigating the observed disparities in network centrality and potentially improving job placement in the faculty hiring market), we designed a network intervention that could be realistically implemented, e.g. through a national fellowship program. Our work provides scientific evidence for the potential impact of programs like the NSF REU[6] which provide research opportunities for young scholars at schools lacking large research facilities. We show how applications for similar programs at the Ph.D. level could be restricted in a principled way to certain demographic groups, certain institutions, or other social characteristics, like parental education, in order to maximize impact. Whether or not such a program is implemented is beyond the scope of our scientific results which clearly demonstrate the need for and potential efficacy of such a program.

Of course, the success of a scientific collaboration depends on the specific people involved, and is more complicated than we assume in our network intervention. At the same time, our intervention was intentionally minimal—adding only a single new coauthorship edge—while many collaborations today involve multiple authors. In reality, a successful publication resulting from such a fellowship program would add a *clique* of size equal to the number of authors on the publication. These many new edges would improve the closeness centrality and predicted placement by more than our simulated intervention predicts. Additional unmeasured benefits of such a program would include the establishment during the fellowship period of new professional relationships beyond the coauthors. That is, we expect that our experiments underestimate the real-world improvements for target scholars.

[6]National Science Foundation Research Experience for Undergraduates: https://www.nsf.gov/funding/initiatives/reu

## 8 Conclusion

The analyses presented here were carried out using a novel data set, hand-collected from computer science department websites in the academic year 2023-24. A de-identified version of this data set is available [Barnes et al. 2025], addressing especially the dearth of demographic information available to previous research, by combining name-based inference and perception to generate demographic meta-labels that align with survey self-reports.

We found that centrality in the computer science coauthorship networks increases with institutional prestige, and we identified demographic disparities in centrality. Ph.D. students and faculty from majority racial groups and with parents employed in tenure or tenure-track positions at Ph.D. granting institutions are more centrally located in the academic collaboration network than their counterparts from minority racial groups without tenure-track parents. By suggesting a single collaboration (edge intervention in the network), we were able to improve the centrality of target individuals with minoritized race identities at lower-ranked institutions. The simulation increases fairness by equalizing social access to information and opportunities across protected groups. This intervention is lightweight and realistic as it does not require access to coauthorship network information, using only institutional prestige and identity data. Furthermore, when applied to Ph.D. student scholars, our simulated intervention improves the predicted rank of targets' placement institutions on the academic job market.

Future work could test whether our proposed intervention works in practice, either using longitudinal data or a real-world program implementation. Such a program may help to reduce observed inequities in the creation and dissemination of scientific knowledge, and experimentally discover the causal factors involved in differential network position and academic career outcomes.

**Generative AI Usage Statement**

The authors acknowledge that generative AI was not used to prepare the manuscript.

## Acknowledgments

This work was supported in part by NSF Awards IIS-1955321 (S.F., M.E.E., N.H. and M.F.), 2219609 (A.C. and K.B.), 2420950 (C.R.) and IIS-1956286 (B.D.S.).

# A Research Methods

## A.1 Coding Procedure

Coders were instructed to search for institutions by name and navigate to the Computer Science department faculty pages. Faculty members' information was collected only if they were listed as assistant professors, associate professors, professors, or distinguished professors. Individuals with other titles such as visiting professor or senior lecturer were not collected, as we assume these faculty are not members of the tenure track. To avoid spelling errors, coders copied and pasted faculty members' names, emails, institutional affiliations, Ph.D. institutions, and titles directly from the webpage. If any of this information was not included on the department page, coders checked faculty members' personal websites. If a faculty member was listed at multiple institutions, coders rectified this ambiguity by referring to the scholar's curriculum vitae, keeping only information for the institution in which they were employed in the present academic year.

Perception data about race and gender was collected based on photos by the hired undergraduate coders. The coding procedure began by searching for faculty photos on department websites. If a photo was not found there, coders were instructed to find another photo by searching, in order of preference, "firstname lastname CurrentInstitution", "firstname lastname linkedin" or "firstname lastname PhDInstitution". Coders ensured these corresponded to the correct individual by verifying that the information presented with the photo, such as titles and institutional affiliations, aligned with what was previously collected. Once an image had been found, coders used faculty photos to categorize scholars based on perceived race and gender. Coders recorded perceived gender as "Man", "Woman", "Non-Binary / Uncertain" or "No photo found". Perceived race was recorded in alignment with U.S. Census Bureau categories as "White", "Black", "Latinx", "East Asian", "Southeast Asian", "South Asian (Indian / Indian subcontinent)", "Middle Eastern / North African", "Native American / Other Indigenous", "Native Hawaiian or Other Pacific Islander", "Multiracial or unsure" or "No photo found" [Bureau 2024]. Coders recorded their own name in a "perceived by" column. Based on self-identification of these seven coders, six are women, one is non-binary, one is White, two are Middle Eastern / North African, two are Hispanic, one is South Asian / Indian, and one is East Asian. While we did not have a large enough sample of student coders to analyze the differential impact of personal gender and racial identification on perception, we note these demographic characteristics since such an impact may exist.

This coding procedure refines a similar one developed in previous research [Clauset et al. 2015]. The full process was discussed in person with the hired coders as a group to ensure consistency and clarify any confusions. Coders were also given the following text documents describing the detailed procedure:

(1) Find the institution's CS faculty page. Google the institution and navigate to the CS department website listing all faculty.
(2) For each faculty member:
    (a) Check the faculty member's title.
        (i) Continue to the next steps for the following titles:
            - Assistant Professor
            - Associate Professor
            - Professor
            - Distinguished Professor
        (ii) Do NOT collect for the following titles:
            - Visiting Assistant Professor
            - Lecturer
            - Senior Lecturer
            - Teaching Assistant Professor
            - Teaching Associate Professor
            - Teaching Professor
            - Research Assistant Professor
            - Research Associate Professor
            - Research Full Professor
            - Emeritus Professor

- Affiliated Faculty
- Professor of the Practice
- Assistant Professor, Lecturer
- Adjunct Faculty

(b) Check the faculty is a professor in Computer Science.
  (i) Do not include affiliated faculty that are not directly Computer Science Professors.

(c) Copy the faculty member's name.

(d) Check if the person is already included in the spreadsheet
  (i) Command + F the name with different variations:
    (A) If middle initial:
      - With the middle initial and without.
  (ii) If they are included but listed at a different institution:
    (A) Google search [Firstname lastname DifferentInstitution]. Look for a personal homepage or department page
    (B) In the following order, check if:
      - The person indicates they moved to the new institution
      - CV includes previous appointments, including both their appointment at new institution and different institution.
      - Photo included matches the photo on the new institution faculty page.
        – If any of the above checks passes: Add the current institution in the "UpdatedInstitution" column.
    (C) Include their title in the "title" column in the spreadsheet.
    (D) If the title is XXX name in front of any of the titles listed in step ai, mark "Yes" in the "Named Position" column.
  (iii) If they are not included:
    (A) Add a row to the spreadsheet (Double click, add row below, to group together with colleagues)
    (B) Do NOT add anything in the "CensusPersonID" column. This is left intentionally blank.
    (C) Paste their name into the "name" column. Note: it's important to use copy/paste here and below to avoid spelling errors.
    (D) Copy and paste the person's email address into the email column
      - If this is not easily accessible / not listed on the site, leave the column blank.
    (E) Copy and paste their current Institution name in the corresponding column.
    (F) Copy and paste their PhD institution in the PhD Institution column
    (G) Include their title in the "title" column in the spreadsheet.
    (H) If the title is XXX name in front of any of the titles listed in step ai, mark "Yes" in the "Name Posiiton" column.
    (I) If any of the above are not included on the main department page, check the faculty member's homepage / personal website.

(e) Perceived race and gender:
  (i) Find a photo of the person. Check if a photo is included on the website found in step 1.
    (A) If the department page does not include a photo, try other sites (but make sure you're getting the right person). Add the website where the photo is found to the spreadsheet in the "notes" column. In order of preference the websites and queries (indicated by square brackets, where university, firstname, etc are substituted by the relevant information for the person) to try are:
      - Firstname lastname CurrentUniversity
      - Firstname lastname linkedin
        – Double check it's the right person by checking the university affiliation

- Firstname lastname PhDInstitution

(ii) Perceived gender. Use your impression of the person based on their photo to fill in the perceived gender column. Options are:
- Man
- Woman
- Non-binary/Uncertain
- No photo found

(iii) Perceived race. Use your impression of the person based on their photo to fill in the perceived race column. Options are:
- White
- Black
- Latinx
- East Asian
- Southeast Asian
- Indian / Indian subcontinent
- Middle Eastern / North African
- Native American / other Indigenous
- Native Hawaiian or Other Pacific Islander
- Multiracial or unsure
- No photo found

(iv) Code pronouns. In the website you found in step 2 or c1, see if the individual refers to themselves in the third person or otherwise indicates their own pronouns.
  (A) If a bio is not included in the website you found in step b or c1, but a personal website is linked, see if the individual refers to themselves in third person in their personal website.
  (B) Do NOT google search for additional/external articles not written by the faculty member.
  (C) Code pronouns those in the pronouns column. Options are:
    - he/him
    - she/her
    - they/them
    - he/them or they/he
    - she/them or they/she
    - Neo pronouns or other
    - No pronouns found

(v) Include your name in the "perceived by" column.

(vi) (OPTIONAL) Add any notes about the data as you enter it.

(vii) If the faculty member is noted as deceased/retired (or Professor Emeritus) on the website you found in step 2 or c1, mark "yes" in the "in the original dataset but not to collect" column.

(viii) If faculty member is noted as left academia / no longer a Professor in CS, mark "yes" in the "in the original dataset but not to collect" column.

In addition to the above procedure, coders matched faculty to DBLP IDs. Where possible, DBLP IDs were automatically matched to faculty in the census by searching first initial, last name, and institution on DBLP and CSrankings. If faculty members had multiple names listed in the DBLP-aliases file, publicly available through the CSrankings repository on GitHub[7], all corresponding names were checked. Exact matches to faculty members' names or CSRankings aliases were recorded, and close matches were checked manually before recording. For faculty with no matches according to this method, coders manually found publication lists on faculty members' institutional websites. Then, coders chose a publication from this list and searched it by name in the publicly available "dblp.xml" file. This procedure resulted in complete coverage, including 2,124 CS faculty with multiple DBLP IDs. The full DBLP ID matching procedure is given below:

- Before the next step:
  1. First Name initial , Last Name, and institution match with CSrankings
  2. First Name initial , Last Name match with CSrankings
  3. First Name initial , Last Name match with DBLP
- Only run for the professors whose matches could not be found automatically using csrankings or dblp.xml.
  1. (only do once) Create a google spreadsheet with the following fields:
     - name
     - institution
     - website
     - dblp_id
  2. Copy and paste the professor name and the institution name into Google
     (a) Go to their institution website
     (b) Add the institution website to your spreadsheet
     (c) Find their publications list
     (d) Copy and paste a publication title into a search query for dblp.xml
        (i) (only do once) download dblp.xml from https://dblp.org/
        (ii) in terminal, open the dblp.xml using: less dblp.xml
        (iii) within less, to open a search query type: /
        (iv) paste the publication name after you type / and press enter
        (v) wait a long time
     (e) Copy and paste the author name that matches the person you're looking for from that publication list of authors - this is the professor's dblp id - into the google spreadsheet
     (f) If multiple publication names are not found in the dblp.xml for a professor, alternatively, go to dblp.org, check for the professor's name (verify it is the right professor using one of their publications on their website) or for one of their publications, and copy the dblp id into the google spreadsheet.
     (g) If publication are not linked in their faculty homepage, check their google scholars webpage (the one verified with the university we have)

[7] https://github.com/emeryberger/CSrankings/blob/gh-pages/dblp-aliases.csv

### A.2 Error analysis of CS Faculty Census

An error analysis was conducted after the completion of the census. We randomly sampled 400 faculty, and re-collected information for these individuals according to the same coding procedure. Then, we compared with the previous collection, noting errors. This analysis revealed 11/400 faculty who should not have been collected because their professional titles (Teaching Professor, Adjunct Professor, Emeritus, etc.) were not in our list for inclusion, an error rate of 2.75%. Additionally, 4 individuals' current or Ph.D. institutions were incorrectly collected, an error rate of 1%.

To ensure inter-coder reliability in the coder-recorded demographics, the perceived race and gender of 283 randomly sampled faculty in our census was double-collected by different individuals. Subsequent analysis showed 89% agreement (251/283) between coders on perceived race and 100% agreement (282/283) on perceived gender.

### A.3 University Prestige Rankings

CSRankings and USNWR ranks were collected from their corresponding websites in November 2024, and are used to corroborate our findings according to the placement power measure of institutional prestige, presented in the main text. Hand-collected university names from our data were linked with the rankings using string matching techniques. Only seven universities from our list were not listed in the placement power ranking: Oakland University, DePaul University, Michigan Technological University, New Mexico Tech, University of Colorado at Colorado Springs, University of Massachusetts Boston, University of Nebraska-Omaha. Eleven universities from our list were not found in CSRankings: University of Tennessee Chattanooga, Catholic University of America, Claremont Graduate University, Santa Clara University, Southern Illinois University-Carbondale, University of Toledo, University of Colorado Denver, University of Maine, University of Mississippi, University of Nevada, Wright State University. Two universities from our list were not found in USNWR: University of Tennessee Chattanooga, University of Wisconsin Milwaukee. No universities were covered by none of these rankings.

We additionally annotated the prestige of faculty members' Ph.D. departments in order to test the importance of Ph.D. prestige in scholars' placement outcomes on the academic job market (Section 6.3). This was done for 2,041 *early career scholars* in our data whose first publication was 2010 or later. These individuals span 179 unique doctoral institutions, of which only one (Oregon Graduate Institute) could not be matched to the "placement power" prestige ranking [Wapman et al. 2022] .

The placement power ranking is calibrated using only data from U.S. Ph.D.-granting institutions, but an additional 223 early career scholars in our census received their Ph.D. degrees from international institutions. These individuals are included in an expanded early career cohort of 2264 individuals which is analyzed using CSRankings and USNWR rankings in Appendices B and C.

### A.4 Name-based inference of demographic categories

We employed four name-based gender (GenderGuesser [Argamon et al. 2003], NamesOrFullNames [Namesor n.d.], NonQuamGender [Buskirk et al. 2023] and WikiGendersort [Bérubé et al. 2020]) and five name-based race inference tools (EthnicolrWiki, EthnicolrCensus, EthnicolrFlorida [Laohaprapanon et al. 2022], Ethnea [Torvik and Agarwal 2016], and EthnicSeer [Treeratpituk and Giles 2012]). These inference tools assign demographic labels based on names alone, a method which is importantly fraught. Names, simply strings of letters, do not have gender or race but are imbued with demographic signal through cultural consensus [Buskirk et al. 2023]. Thus, the association between a name and a particular race or gender is not fixed, but varies by culture and over time [Haslanger 2000]. Some name-based algorithms seek to address this by including data from across time periods and allowing users to specify time as a variable.

The name-based inference tools we used report the uncertainty of the association between names and demographics via a probability value, indicating the strength of the gender or race signal of a particular name. The demographic labels and their associated probabilities are data-driven estimates. For example, NonQuam [Buskirk et al. 2023] labels the name "Sally" as gendered female with $p = 0.984$, indicating that on average 98.4% of occurrences of the name "Sally" in sources from NonQuam's database are gendered female.

A major limitation of tools for name-based gender inference is that they exclusively return binary gender categories [Scheuerman and Brubaker 2018]. Therefore, name-based inference tools could not correctly label any non-binary faculty in our census. Additionally, racial categories returned by name-based inference methods did not always align with the U.S. census categories. Thus, we manually aligned the race labels from name-based inference with our categories. EthnicolrFlorida was only used to label Latinx, Black, and White individuals due to reporting only one Asian category which did not match ours and returning no other categories. EthnicolrCensus was not used due to reporting only four, broad racial categories and performing poorly in alignment with survey self-reports. Ethnea, EthnicolrWiki and Ethnicseer returned many racial categories. Below are details showing how race categories from the name-based inference methods we employed were matched to our expanded U.S. census race categories.

**EthnicolrWiki:**

- White: GreaterEuropean (British), GreaterEuropean (EastEuropean), GreaterEuropean (Jewish), GreaterEuropean (WestEuropean, French), GreaterEuropean (WestEuropean, Germanic), GreaterEuropean (WestEuropean, Italian), GreaterEuropean (WestEuropean, Nordic)
- East Asian: Asian (GreaterEastAsian, EastAsian), Asian (GreaterEastAsian, Japanese)
- South Asian (Indian / Indian subcontinent): Asian (IndianSubContinent)
- Middle Eastern / North African: GreaterAfrican (Muslim)
- Latinx: GreaterEuropean (WestEuropean, Hispanic)
- Black: GreaterAfrican (Africans)
- Native Hawaiian or Pacific Islander: None
- Southeast Asian: None

**Ethnicseer:**

- White: eng, ger, frn, ita, rus
- East Asian: chi, jap, kor
- South Asian (Indian / Indian subcontinent): ind
- Middle Eastern / North African: mea
- Latinx: spa
- Black: None
- Native Hawaiian or Pacific Islander: None
- Southeast Asian: vie

**Ethnea:**

- White: ENGLISH, GERMAN, FRENCH, ITALIAN, SLAV, NORDIC, DUTCH, HUNGARIAN, ROMANIAN, ISRAELI
- East Asian: CHINESE, JAPANESE, KOREAN, MONGOLIAN
- South Asian (Indian / Indian subcontinent): INDIAN
- Middle Eastern / North African: ARAB, TURKISH, GREEK
- Latinx: HISPANIC, CARIBBEAN
- Black: AFRICAN
- Native Hawaiian or Pacific Islander: POLYNESIAN
- Southeast Asian: VIETNAMESE, THAI, INDONESIAN

### A.5 Survey text

Our survey, approved by the Haverford College IRB dated June 17, 2024 under the title "Demographics and Faculty Co-Authorship Networks.", was sent to all faculty in our census in the summer of 2024 using the emails collected as part of the data coding procedure. Survey respondents provided their name, confirmed they were currently employed as tenured or tenure-track faculty, and selected a single gender attribute: male, female, non-binary or other. Respondents were also asked to identify their race/ethnicity by choosing one or multiple racial categories from the same list of options provided to coders. For both race and gender questions, respondents were provided an optional free-form text box to provide further information if desired. Fifty-five individuals used this space to report other racial identities, predominantly to identify as Jewish, and 11 individuals self-described their gender. The complete text of the survey questionnaire is provided below:

This survey asks you about your demographics and parental educational history and takes about 1 minute to complete.

The data will be used to inform an analysis of the influence of faculty demographics and socioeconomic status on advancement, productivity, and coauthorship patterns. The goal of this research is to support equity by better understanding how demographics relate to and shape faculty experiences and connections such as coauthorship relationships in faculty networks, with the broader aim of promoting equal access to information for all groups, especially those currently underrepresented within computer science.

Your responses will be used for the purposes of this study and the data may also be used for future research also focusing on demographics and socioeconomic status and faculty coauthorship networks. Such research could occur indefinitely in the future, and individuals will not be alerted to each new publication.

The study "Demographics and Faculty Co-Authorship Networks" (IRB approved) is carried out by researchers from Haverford College. This study is supported by the National Science Foundation (NSF IIS-1955321).

Your participation is voluntary and you may stop at any time. Your information will be handled confidentially, but can not be truly anonymous given its association with a coauthorship network, and there is a risk of data leakage. The specific data stored about you indefinitely will include: name, title, email, institution, PhD institution, gender, race, parental degree attainment, parental faculty status, and publications with coauthorship information. The self-reported information from this survey will not be released publicly; any published information will be based on aggregated data.

If you have further questions about the research or your rights as a research participant, please contact Prof. Sorelle Friedler via sorelle@cs.haverford.edu. You may also address any concerns to the chairperson of Haverford College's IRB (a committee with oversight over human subject research) via hc-irb@haverford.edu.

To proceed, please check the box next to the following statement:

- I have read and understood the consent form and give my permission to participate in this study.

(1) Please provide the full name under which you publish academic work.

(2) Are you currently employed as a tenured or tenure-track faculty member in a department that grants PhDs in computer science?
  - Yes
  - No

(3) How do you identify your gender?
  - Male
  - Female
  - Non-Binary
  - Other
  - Free-form text box

(4) How do you identify your race / ethnicity? (Select one or more)
  - White
  - Black
  - Latinx
  - East Asian
  - Southeast Asian
  - South Asian (Indian / Indian subcontinent)
  - Middle Eastern / North African
  - Native American or other Indigenous
  - Native Hawaiian or Pacific Islander
  - Other

- Free-form text box

(5) What is the highest educational attainment of any of your parents?
- Some high school
- High school
- Some college
- College degree
- Masters degree
- Doctorate (professional, e.g., JD, MD)
- Doctorate (research, e.g., Ph.D.)

(6) Have any of your parents ever been a tenured or tenure-track professor at a PhD- granting institution?
- Yes
- No

## A.6 Demographic Meta-Labeling

A general schematic for the race and gender meta-labelers described above is given in Algorithm 1. To determine the resulting gender and race meta-labelers, we search through a space of many algorithms determined by the specific choices of for loop orderings (Line 5) and thresholds (Line 7). We describe this process and results for the final race and gender meta-labelers next.

**Algorithm 1** Demographic Meta-Labeler

```
Input: NameMap : name → ( inferenceMethods, perceptionLabel),
       inferenceMethods : demographicLabel → probability,
       allDemographicLabels
Output: ResultMap: name → metaLabel

1:  ResultMap ⟵ ∅ // Initialize all MetaLabels as None.
2:  for name ∈ NameMap do
3:      labeled ⟵ False
4:      inferenceMethods, perceptionLabel ⟵ NameMap[name]
5:      for L ∈ allDemographicLabels do
6:          for inference ∈ inferenceMethods do
7:              if inference[L] ≥ threshold and ¬labeled then
8:                  ResultMap[Name] ⟵ L
9:                  labeled ⟵ True
10:             end if
11:         end for
12:     end for
13:     if ¬labeled then
14:         ResultMap[Name] ⟵ perceptionLabel
15:     end if
16: end for
```

## A.7 Race Meta-Labeling

*Brute force methodology for finding race meta-labeler.* In determining the specific meta-labeling algorithm to use for racial inference, following the general pattern described in Algorithm 1, we must determine the order in which to apply racial labels (the order of iteration for Line 5). Using a brute force search, we examine all permutations of the racial labels used throughout our census, and choose the label ordering that resulted in the meta-labeler with the best alignment with the survey data, measured as accuracy. The chosen order is critical to the algorithm since it can resolve false positives. A name misclassified as White in one ordering could be correctly classified in a different ordering which labels the name as belonging to the self-reported racial category before applying any White labels.

For each name-based inference tool, we selected separate confidence thresholds for every race by using the lowest threshold that still minimized false positives. For example, names of individuals who self-reported as Black were more commonly mislabeled English than White, and names labeled as Jewish by the inference tools were self-reported as Jewish, White or Middle Eastern. Thus, we selected confidence thresholds such that no individuals who self-reported as Black were labeled English by Ethnea ($p \geq 0.99$). EthniclrWiki mislabeled one individual who self-reported Black with $p = 0.99$, but a threshold of $p \geq 0.91$, captured 73 individuals with just this one false positive.

With these inference algorithms and associated thresholds chosen, the brute force search of permutations of eight racial categories (White, Indian, East Asian, Middle Eastern / North African, Black, Latinx, English and Jewish) resulted in 137 possible meta-labeling algorithms that tied with an accuracy of 92% (733 / 797) in comparison to the survey data. The accuracy of meta-algorithms is limited by the accuracies of their sub-components, and here the maximum accuracy achievable was 96% (766/797). Most of the algorithms tied with 92% accuracy labeled White or English names first, due to strong alignment with survey self-reports for these categories. From among these top-performing permutations, we chose an algorithm which instead labeled non-White names first to reduce false positives in which an individual who does not identify as White is labeled as White. Pseudocode for the final race meta-labeler is given in Algorithm 2. The ordering of race categories, thresholds for each race and name-based inference method applied in Algorithm 2 are given in Table 1.

Applying Algorithm 2 to 797 respondents who answered survey race questions resulted in 52 people classified with perception. Results for individuals who self-reported one race are shown in the confusion matrix in Figure 1. The race meta-labeler identifies White, Indian and East Asian names well, achieving group accuracies over 90% in all cases. and performs worse for minority groups (Figure 1) often misidentifying Middle Eastern / North African or Black names as White. Many of the people incorrectly predicted as White who self-identify as Middle Eastern were due to names labeled as Jewish by Ethnea or EthnicolrWiki. These were individuals who wrote "Jewish" in the provided text box, but in some cases chose Middle Eastern / North African and in other cases White from our check list. We chose to label Ethnea and EthnicolrWiki Jewish names as White, but in many cases Middle Eastern / North African would also be appropriate. Out of the 55 respondents who reported multiple racial categories, our algorithm correctly classified them with one of their reported races in all but 2 cases. Overall, our accuracies are comparable to the popular BISG[8] method without requiring address information.

*Exhaustive methodology.* The brute force method described above was corroborated by comparison to an exhaustive search for the best algorithm. This methodology at every step checks the accuracy of every name-based inference tool (checking all ethnicities and all thresholds) and chooses the classification which results in the least false positives. Using this method, 156 people can be classified with 0 false positives. 646 more people are classified automatically by allowing for either 0 or 1 false positives. Any further automatic classifications required accepting 2 or more false positives and performed worse than applying perception labels to the same names. Therefore, the remaining 113 people were classified with perception. This exhaustive search resulted in an overall accuracy of 734/797 (92%). Thus, the results of the algorithm found using the exhaustive search approach are comparable to Algorithm 2 found by brute force, indicating that we are achieving the maximum accuracy possible given our survey data set.

[8] https://www.rand.org/health/surveys/tools/bisg.html

**Algorithm 2** Race Meta-Labeler

**Input**: NameMap: → ( Ethnea → ( raceLabel, probability ), EthnicolrWiki → ( raceLabel, probability ), EthnicolrFlorida → ( raceLabel, probability ), perceptionLabel )
**Output**: ResultMap: name → metaLabel

```
1:  for name ∈ NameMap do
2:      labeled ← False
3:      Ethnea, EthnicolrWiki, EthnicolrFlorida, perceptionLabel ← NameMap[name]
4:      if Ethnea[“RaceLabel”] == “Southeast Asian” and Ethnea[“p”] ≥ 0.90 then
5:          ResultMap[Name] ← “Southeast Asian”
6:          labeled ← True
7:      else if (Ethnea[“RaceLabel”] == “Indian / Indian subcontinent” and Ethnea[“p”] ≥ 0.67) then
8:          ResultMap[Name] ← “Indian / Indian subcontinent”
9:          labeled ← True
10:     else if (EthnicolrWiki[“RaceLabel”] == “Indian / Indian subcontinent” and EthnicolrWiki[“p”] ≥ 0.57)
    then
11:         ResultMap[Name] ← “Indian / Indian subcontinent”
12:         labeled ← True
13:     else if (Ethnea[“RaceLabel”] == “Jewish” and Ethnea[“p”] ≥ 0.45) then
14:         ResultMap[Name] ← “White”
15:         labeled ← True
16:     else if (EthnicolrWiki[“RaceLabel”] == “Jewish” and EthnicolrWiki[“p”] ≥ 0.61) then
17:         ResultMap[Name] ← “White”
18:         labeled ← True
19:     else if (Ethnea[“RaceLabel”] == “English” and Ethnea[“p”] ≥ 0.99) then
20:         ResultMap[Name] ← “White”
21:         labeled ← True
22:     else if (EthnicolrWiki[“RaceLabel”] == “English” and EthnicolrWiki[“p”] ≥ 0.91) then
23:         ResultMap[Name] ← “White”
24:         labeled ← True
25:     else if (Ethnea[“RaceLabel”] == “East Asian” and Ethnea[“p”] ≥ 0.69) then
26:         ResultMap[Name] ← “East Asian”
27:         labeled ← True
28:     else if (EthnicolrWiki[“RaceLabel”] == “East Asian” and EthnicolrWiki[“p”] ≥ 0.66) then
29:         ResultMap[Name] ← “East Asian”
30:         labeled ← True
31:     else if (Ethnea[“RaceLabel”] == “White” and Ethnea[“p”] ≥ 0.74) then
32:         ResultMap[Name] ← “White”
33:         labeled ← True
34:     else if (EthnicolrWiki[“RaceLabel”] == “English” and EthnicolrWiki[“p”] ≥ 0.74) then
35:         ResultMap[Name] ← “White”
36:         labeled ← True
37:     else if (EthnicolrFlorida[“RaceLabel”] == “White” and EthnicolrWiki[“p”] ≥ 0.97) then
38:         ResultMap[Name] ← “White”
39:         labeled ← True
40:     else if (Ethnea[“RaceLabel”] == “Middle Eastern / North African” and Ethnea[“p”] ≥ 0.99) then
41:         ResultMap[Name] ← “Middle Eastern / North African”
42:         labeled ← True
43:     else if (EthnicolrWiki[“RaceLabel”] == “Middle Eastern / North African” and EthnicolrWiki[“p”] ≥
    0.77) then
44:         ResultMap[Name] ← “Middle Eastern / North African”
45:         labeled ← True
```

| Race | Order | Ethnea | EthnicolrWiki | EthnicolrFlorida |
|---|---|---|---|---|
| Southeast Asian | 1st | 0.90 | - | - |
| South Asian | 2nd | 0.67 | 0.57 | - |
| Jewish | 3rd | 0.45 | 0.61 | - |
| English | 4th | 0.99 | 0.91 | - |
| East Asian | 5th | 0.69 | 0.66 | - |
| White | 6th | 0.74 | 0.74 | 0.97 |
| MidEast/NorAfr | 7th | 0.99 | 0.77 | - |
| Latinx | 8th | 0.95 | 0.72 | 0.90 |
| Black | 9th | - | - | 0.77 |

Table 1. Thresholds and ordering of race categories in meta-labeling algorithm.

**Algorithm 2** Race Meta-Labeler

```
    else if (Ethnea[“RaceLabel”] == “Latinx” and Ethnea[“p”] ≥ 0.95) then
        ResultMap[Name] ← “Latinx”
        labeled ← True
    else if (EthnicolrWiki[“RaceLabel”] == “Latinx” and EthnicolrWiki[“p”] ≥ 0.72) then
        ResultMap[Name] ← “Latinx”
        labeled ← True
    else if (EthnicolrFlorida[“RaceLabel”] == “Latinx” and EthnicolrFlorida[“p”] ≥ 0.90) then
        ResultMap[Name] ← “Latinx”
        labeled ← True
    else if (EthnicolrFlorida[“RaceLabel”] == “Black” and EthnicolrFlorida[“p”] ≥ 0.77) then
        ResultMap[Name] ← “Black”
        labeled ← True
    end if
    if ¬labeled then
        ResultMap[Name] ← perceptionLabel
    end if
end for
```

Algorithm 2 found by the brute force method described in the text is preferable to the algorithm found by exhaustive search because it requires classifying less people using perception and it is less likely to overfit the survey data. This is evident by the results of a 5-fold cross-validation. The mean training accuracy for 5 folds was 91.97% for the exhaustive search method and 92.28% for the brute force method, whereas their mean test accuracies were 89.59% and 92.22% respectively.

## A.8 Gender Meta-Labeling

For the gender meta-labeler, the key search space for Algorithm 1 is over possible thresholds per inference method (Line 7); we search for the thresholds which maximizes the accuracy of the resulting label with respect to survey data. Our identified thresholds result in zero individuals who self-reported as men or women mislabeled by name-based inference.

NonQuamGender [Buskirk et al. 2023] achieves 87% accuracy overall and 98% accuracy on faculty names which were self-reported as men or women. A threshold of 0.75 applied to all faculty names in our survey data maximized alignment between NonQuam and self-identified labels, resulting in only six mislabeled self-identified men or women. All six of these individuals received East Asian as their top probability race label according to Ethnea. A NonQuam threshold of 0.85 for East Asian names captured 4 of these misclassified individuals. Two were still labeled incorrectly due to NonQuam gender signal below our threshold and a "no photo found" label from perception coders.

Applying the gender meta-labeler to 811 respondents who answered survey gender questions resulted in 117 people classified with perception. The algorithm did not misclassify any self-reported women and misclassified two self-reported men because no photo was found for perception labeling. Name-based inference applies binary gender labels, therefore our algorithm can only label individuals as non-binary in the perception step. Although 3 out of the 15 self-identified non-binary scholars were labeled as non-binary by coders, in our algorithm these names are first mislabeled as men or women by NonQuam, resulting in a misclassification of all 15 self-reported non-binary scholars. Overall, Algorithm 3 achieves an accuracy of 98% (794/811) on survey data, as compared to a maximum possible accuracy of 99% (800/811).

Algorithm 3 was used to label the gender of all 5,670 faculty in our census. The NonQuamGender thresholds used are given in Table 2.

| Gender | East Asian | Order | NonQuamGender Threshold |
|---|---|---|---|
| Woman | Yes | 1st | 0.85 |
| Man | Yes | 2nd | 0.85 |
| Woman | No | 3rd | 0.75 |
| Man | No | 4th | 0.75 |

Table 2. Thresholds for gender meta-labeling algorithm.

### A.9 Comparisons with the Taulbee Reports

Our gender labels for the total population fall within 5% of Taulbee estimates for every category (full, associate, and assistant professors), and our race labels for the total population are within 10% of Taulbee estimates. Typically, Taulbee reported lower percentages of non-White faculty than we found, so we expect the difference is in part due to Taulbee's inclusion of "non-resident alien" and "residency unknown" categories which our label set does not include. In Tables 3–6 below, the Taulbee report's "other" column includes teaching professors, other instructors, researchers and postdocs.

## B Network description

Figure 7 shows correlation between closeness centrality and university rank according to CSRankings and USNWR.

Figure 8 shows demographic and prestige disparities in degree and closeness centrality.

**Algorithm 3** Gender Meta-Labeler

**Input**: NameMap: → ( NonQuam → ( genderLabel, probability ), Ethnea → ( raceLabel, probability ), perceptionLabel )

**Output**: ResultMap: name → metaLabel

```
1: // Initialize all MetaLabels as None.
2: for name ∈ NameMap do
3:     labeled ← False
4:     NonQuam, Ethnea, perceptionLabel ← NameMap[name]
5:     if Ethnea["RaceLabel"] == "East Asian" then
6:         if (NonQuam["GenderLabel"] == "Female" and NonQuam["p"] ≥ 0.85) then
7:             ResultMap[Name] ← "Woman"
8:             labeled ← True
9:         else if (NonQuam["GenderLabel"] == "Male" and NonQuam["p"] ≥ 0.85) then
10:            ResultMap[Name] ← "Man"
11:            labeled ← True
12:        end if
13:    else
14:        if (NonQuam["GenderLabel"] == "Female" and NonQuam["p"] ≥ 0.75) then
15:            ResultMap[Name] ← "Woman"
16:            labeled ← True
17:        else if (NonQuam["GenderLabel"] == "Male" and NonQuam["p"] ≥ 0.75) then
18:            ResultMap[Name] ← "Man"
19:            labeled ← True
20:        end if
21:    end if
22:    if ¬labeled then
23:        ResultMap[Name] ← perceptionLabel
24:    end if
25: end for
```

| | Full | Associate | Assistant | Other | Total |
|---|---|---|---|---|---|
| Male | 2049 (81%) | 1108 (78%) | 1257 (72%) | 2207 (72%) | 6621 (76%) |
| Female | 455 (18%) | 315 (22%) | 501 (28%) | 865 (28%) | 2136 (24%) |
| Non-binary | 9 (1%) | 0 (0%) | 3 (0%) | 6 (0%) | 18 (0%) |
| Unknown | 215 | 84 | 119 | 185 | 603 |
| Total | 2728 | 1507 | 1880 | 3263 | 9378 |

Table 3. Gender of current US faculty as reported in the 2024 Taulbee report (Table F6).

| | Full | Associate | Assistant | Distinguished | Total (n=5670) |
|---|---|---|---|---|---|
| Male | 2043 (84%) | 1112 (80%) | 1217 (75%) | 178 (82%) | 4550 (80%) |
| Female | 388 (16%) | 273 (20%) | 406 (25%) | 40 (18%) | 1107 (20%) |
| Non-binary | 1 | 3 | 4 | 0 | 8 |
| No photo | 1 | 2 | 2 | 0 | 5 |
| Total | 2433 | 1390 | 1629 | 218 | 5670 |

Table 4. Gender of current US faculty in our data, replicating the Taulbee calculations.

| | Full | Associate | Assistant | Other | Total |
|---|---|---|---|---|---|
| Nonresident Alien | 56 (2%) | 39 (3%) | 269 (17%) | 290 (10%) | 654 (8%) |
| American Native | 34 (2%) | 5 (1%) | 31 (2%) | 12 (0%) | 82 (1%) |
| Asian | 735 (32%) | 437 (34%) | 616 (38%) | 565 (20%) | 2353 (32%) |
| Black | 28 (1%) | 29 (2%) | 32 (2%) | 79 (0%) | 168 (2%) |
| Pacific Islander | 0 (0%) | 1 (0%) | 2 (0%) | 2 (0%) | 5 (0%) |
| White | 1335 (58%) | 689 (53%) | 538 (33%) | 1582 (57%) | 4144 (52%) |
| Multiracial | 15 (2%) | 11 (1%) | 18 (1%) | 12 (0%) | 56 (0%) |
| Hispanic | 52 (2%) | 37 (3%) | 50 (3%) | 92 (3%) | 231 (2%) |
| Unknown | 57 (3%) | 53 (4%) | 64 (4%) | 138 (5%) | 312 (3%) |
| Total | 2728 | 1507 | 1880 | 2772 | 9378 |

Table 5. Race of current US faculty as reported in the 2024 Taulbee report (Table F7).

| | Full | Associate | Assistant | Distinguished | Total (n=5670) |
|---|---|---|---|---|---|
| South Asian | 409 (17%) | 177 (13%) | 287 (18%) | 44 (20%) | 917 (16%) |
| East Asian | 460 (19%) | 349 (25%) | 585 (36%) | 32 (15%) | 1426 (25%) |
| Southeast Asian | 14 (1%) | 5 (0%) | 13 (1%) | 1 (1%) | 33 (1%) |
| Asian (all) | 883 (36%) | 531 (38%) | 885 (54%) | 77 (35%) | 2376 (42%) |
| Black | 11 (1%) | 18 (2%) | 30 (2%) | 1 (1%) | 60 (1%) |
| MidEast/NorAfr | 143 (6%) | 130 (10%) | 162 (10%) | 9 (4%) | 444 (8%) |
| White | 1320 (55%) | 658 (47%) | 491 (31%) | 126 (57%) | 2595 (46%) |
| Latinx | 58 (2%) | 38 (3%) | 51 (3%) | 4 (2%) | 151 (3%) |
| Multiracial | 13 (0%) | 7 (0%) | 8 (0%) | 1 (1%) | 29 (0%) |
| No photo | 4 (0%) | 6 (0%) | 1 (0%) | 0 (0%) | 11 (0%) |
| Total | 2433 | 1390 | 1629 | 218 | 5670 |

Table 6. Race of current US faculty in our data, replicating the Taulbee calculations.

## C Interventions

Figure 9 shows the results of our basic intervention (Section 6.1) on the closeness centrality and predicted placement of faculty in our census when target and sponsor individuals are defined by their USNWR and CSRankings institutional rank.

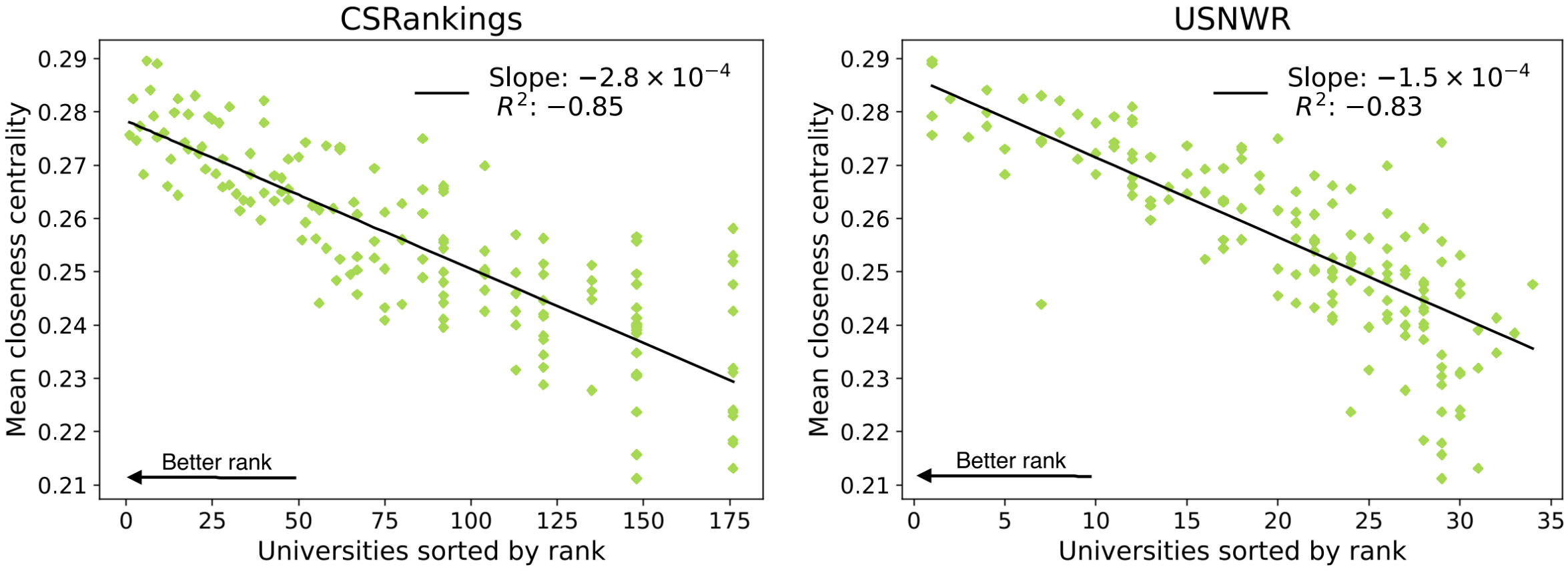

Fig. 7. University prestige (rank, using the measure of placement power [Wapman et al. 2022]; smaller score is more prestigious) versus the mean closeness centrality of faculty at that institution.

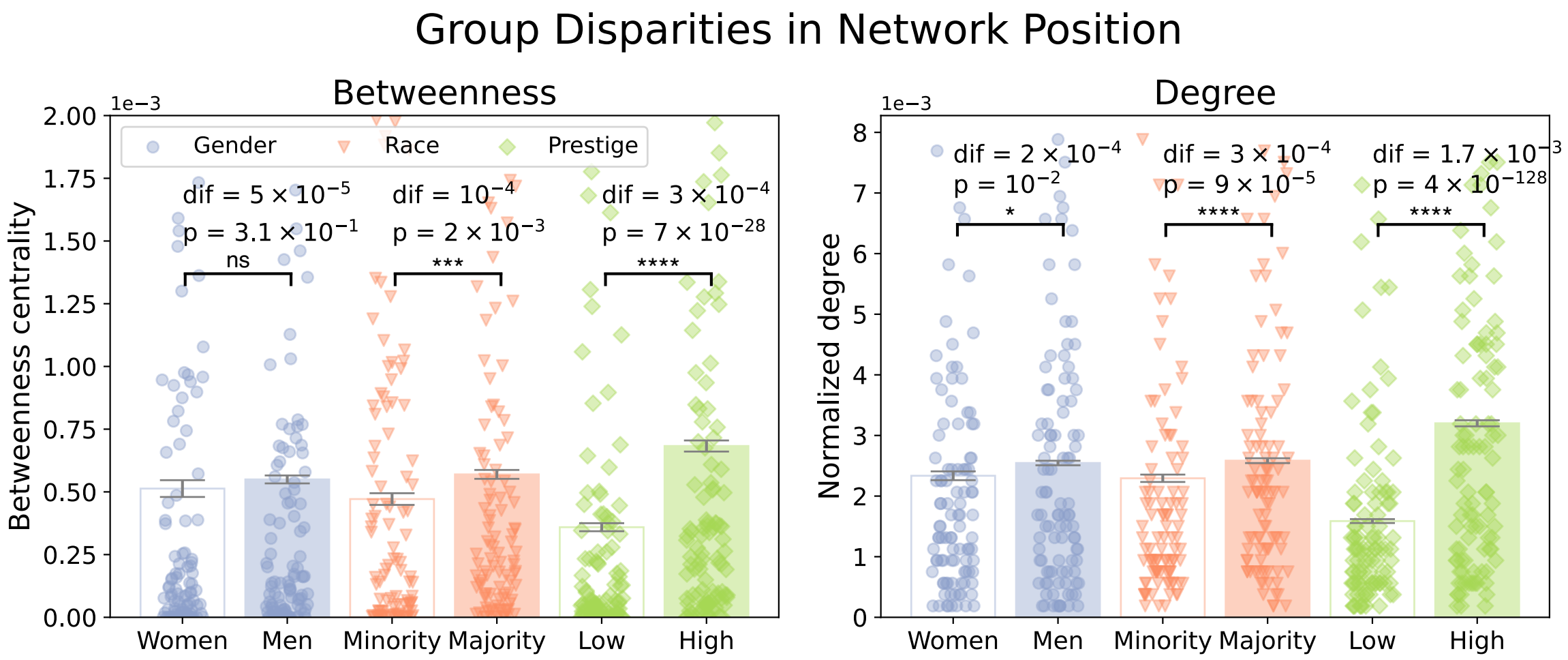

Fig. 8. The p-values reported are from pooled t-tests comparing the betweenness centrality (left) and degree (right) of women and men; majority and minority race groups; and individuals currently employed at universities ranked above or below the median in our data. Standard error of the mean (SEM) error bars.

Figure 10 shows the results of our Ph.D. intervention (Section 6.3) on the closeness centrality and predicted placement of Ph.D. scholars when target and sponsor individuals are defined by their USNWR institutional rank.

Figure 11 shows the results of our Ph.D. intervention (Section 6.3) on the closeness centrality and predicted placement of Ph.D. scholars when target and sponsor individuals are defined by their CSRankings institutional rank.

| Education = 1 Group | Education = 0 Group | Difference in Closeness Centrality | P Value |
|---|---|---|---|
| Research doctorate | Professional doctorate, Masters degree, Other | $4 \times 10^{-3}$ | 0.04 |
| Professional doctorate | Research doctorate, Masters degree, Other | $2 \times 10^{-3}$ | 0.55 |
| Professional doctorate, Research doctorate | Masters degree, Other | $3 \times 10^{-3}$ | 0.08 |
| Professional doctorate, Research doctorate, Masters degree | Other | $2 \times 10^{-3}$ | 0.21 |

Table 7. **Difference in faculty closeness centrality by parental education.** Pooled t-tests comparing faculty responses (n=820) to survey question about parental education. Faculty with more highly educated parents have higher closeness centrality in cumulative collaboration network. "Other" category in table includes "college degree", "some college", "high school" and "some high school" survey responses. Professional doctorate includes e.g. J.D. or M.D. degrees. Research doctorate includes e.g. Ph.D. degrees.

| Education = 1 Group | Education = 0 Group | Difference in Closeness Centrality | P Value |
|---|---|---|---|
| Research doctorate | Professional doctorate, Masters degree, Other | $-2 \times 10^{-5}$ | 0.99 |
| Professional doctorate | Research doctorate, Masters degree, Other | $4 \times 10^{-3}$ | 0.27 |
| Professional doctorate, Research doctorate | Masters degree, Other | $1 \times 10^{-3}$ | 0.50 |
| Professional doctorate, Research doctorate, Masters degree | Other | $7 \times 10^{-4}$ | 0.82 |

Table 8. **Difference in faculty closeness centrality from Ph.D. networks by parental education.** Pooled t-tests comparing faculty responses (n=310) to survey question about parental education. Positive differences indicate faculty with more highly educated parents have higher closeness centrality. "Other" category in table includes "college degree", "some college", "high school" and "some high school" survey responses. Professional doctorate includes e.g. J.D. or M.D. degrees. Research doctorate includes e.g. Ph.D. degrees.

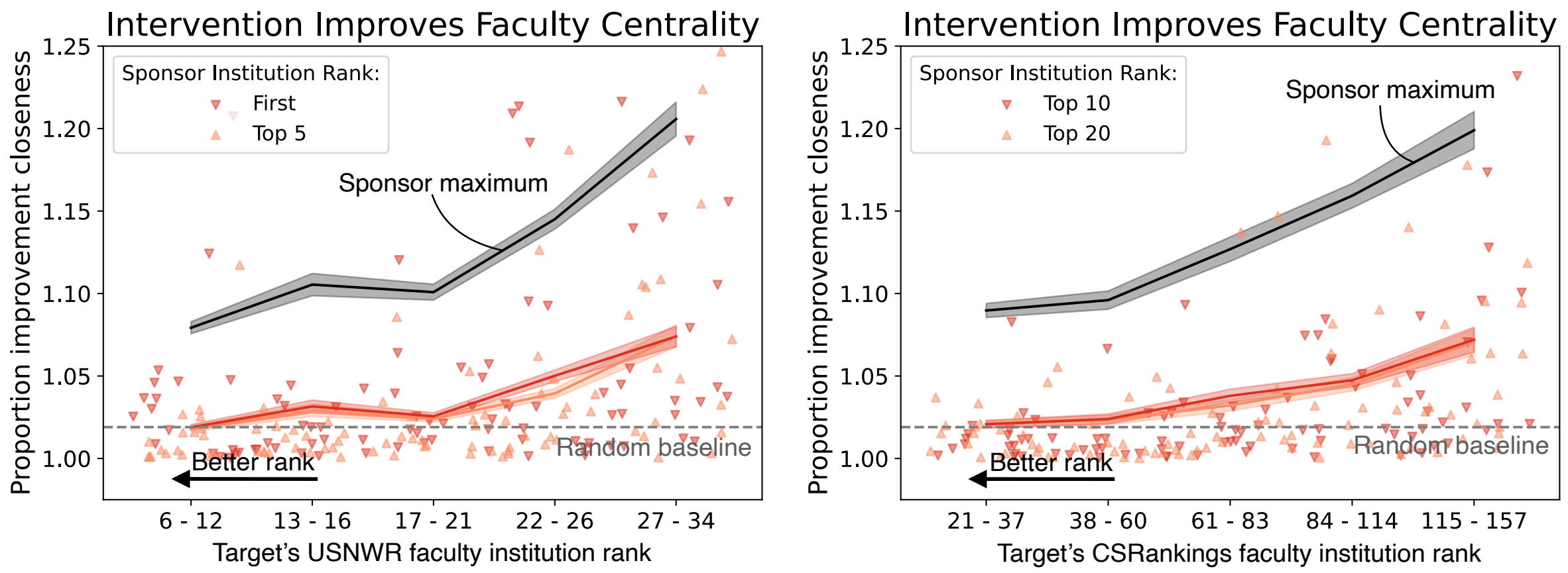

Fig. 9. This simulated intervention results in increases in the closeness centrality of target individuals with minoritized race identities at low prestige current institutions (rank using Wapman et al. [2022] placement power). Our interventions (red lines) are compared to greedy selection of sponsor who would maximally improve target's closeness (black line) and to a random baseline.

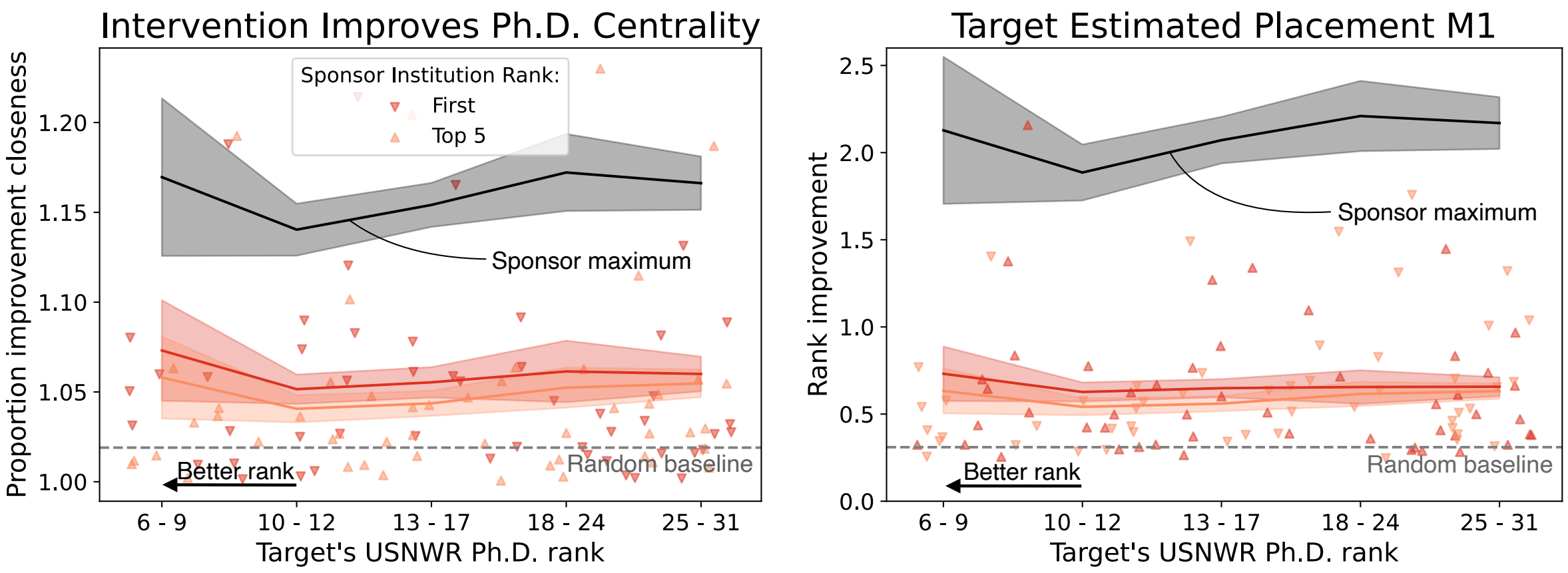

Fig. 10. USNWR: Simulated collaboration increases the centrality of target individuals at low ranked institutions and improves the estimated rank of their placement institutions on the academic job market.

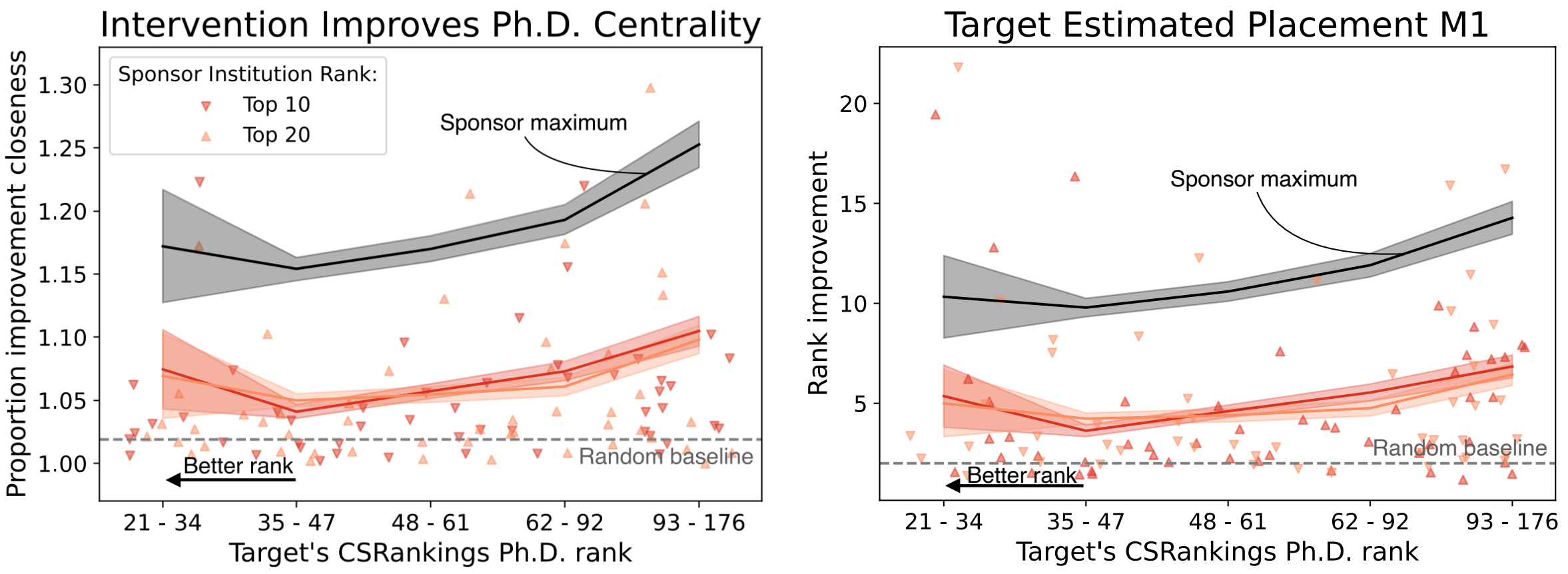

Fig. 11. CSRankings: Simulated collaboration increases the centrality of target individuals at low ranked institutions and improves the estimated rank of their placement institutions on the academic job market.

## C.1 Universities

This study examines data for faculty from 178 Ph.D. granting institutions in the U.S. originally compiled in previous research [Clauset et al. 2015]:

University of Utah, Auburn University, Univ. of Illinois at Urbana-Champaign, University of South Carolina, University of South Florida, California Institute of Technology, Carnegie Mellon University, Indiana University, University of Toledo, New York University, Oakland University, Princeton University, Northwestern University, University of Massachusetts Amherst, University of New Mexico, Arizona State University, University of Texas at Austin, Virginia Commonwealth University, Iowa State University, New Jersey Institute of Technology, Florida State University, Texas A&M University, Boston University, Brandeis University, Brigham Young University, Brown

University, Georgia Institute of Technology, University of Michigan, Univ. of California-Irvine, Univ. of California-Berkeley, Columbia University, Case Western Reserve University, George Mason University, Catholic University of America, Claremont Graduate University, Clarkson University, Clemson University, University of Florida, Florida Atlantic University, College of William and Mary, North Carolina State University, Colorado School of Mines, Colorado State University, Yale University, Cornell University, University of Texas at Dallas, Dartmouth College, DePaul University, Drexel University, University of Maryland-College Park, Duke University, Florida Institute of Technology, Florida International University, University of North Carolina-Charlotte, University of Missouri-Kansas City, Virginia Tech, State University of New York-Albany, George Washington University, University of Colorado Boulder, Georgia State University, University of Texas at San Antonio, Northeastern University, Univ. of California-Santa Barbara, Harvard University, Illinois Institute of Technology, University of Illinois at Chicago, Pennsylvania State University, University of Virginia, Binghamton University, Johns Hopkins University, Rice University, Kansas State University, Kent State University, Lehigh University, Louisiana State University, Missouri University of Technology, Michigan State University, Michigan Technological University, Western Michigan University, University of Cincinnati, Mississippi State University, Stonybrook University, Vanderbilt University, Massachusetts Institute of Technology, Montana State University, University of North Texas, Naval Postgraduate School, University of Southern Mississippi, New Mexico Tech, New Mexico State University, Stanford University, North Dakota State University, Nova Southeastern University, Purdue University, Ohio State University, University of Georgia, Ohio University, Oklahoma State University, Old Dominion University, Oregon Health & Science University, Oregon State University, University of Wisconsin-Madison, Univ. of California-Santa Cruz, Portland State University, Rensselaer Polytechnic Institute, Rochester Institute of Technology, Rutgers University, Santa Clara University, Southern Illinois University-Carbondale, Southern Methodist University, Univ. of California-San Diego, University of Washington, University at Buffalo, Stevens Institute of Technology, Syracuse University, Univ. of California-Riverside, Temple University, Texas Tech University, Tufts University, Univ. of California-Davis, University of Southern California, University of Chicago, Univ. of California-Los Angeles, University of Alabama-Birmingham, University of Tennessee-Chattanooga, University of Alabama, University of Alabama-Huntsville, University of Arizona, University of Arkansas, University of Arkansas-Little Rock, University of Central Florida, University of Colorado-Denver, University of Colorado at Colorado Springs, University of Connecticut, University of Delaware, University of Denver, University of Hawaii at Manoa, University of Houston, University of Idaho-Moscow, University of Pennsylvania, University of Iowa, University of Kansas, University of Massachusetts Lowell, University of Kentucky, Washington University in St. Louis, University of Louisiana-Lafayette, University of Louisville, University of Massachusetts Boston, University of Maine, Univ. of Maryland-Baltimore County, University of Miami, University of Memphis, University of Minnesota, University of Mississippi, University of Missouri, University of Nebraska, University of Texas at Arlington, University of Nebraska-Omaha, University of Nevada Las Vegas, University of Nevada, University of New Hampshire, University of North Carolina, Worcester Polytechnic Institute, University of Notre Dame, University of Oklahoma, University of Oregon, University of Pittsburgh, University of Rhode Island, University of Rochester, Wright State University, University of Tennessee, University of Texas-El Paso, University of Tulsa, University of Wisconsin-Milwaukee, University of Wyoming, Utah State University, Washington State University, Wayne State University